\documentclass[
nofootinbib,
 amsmath,amssymb,
 aps,twocolumn,nofootinbib
]{revtex4-1}
\usepackage[parfill]{parskip}
\usepackage{graphicx}
\usepackage{amssymb}
\usepackage{slashed}
\usepackage{epstopdf}
\usepackage{mathtools}
\usepackage{color}
\usepackage{booktabs}
\usepackage{slashed} 

\newcommand{\be}{\begin{equation}}
\newcommand{\ee}{\end{equation}}
\newcommand{\ba}{\begin{eqnarray}}
\newcommand{\ea}{\end{eqnarray}}

\begin{document}
\title{The lightest flavor--singlet $qqq$ baryons as   witnesses to color}
\author{Jonathan Est\'evez, Felipe J. Llanes--Estrada}
\affiliation{Universidad Complutense de Madrid, Departamento de F\'isica Te\'orica and IPARCOS.}
\author{ V\'{i}ctor Mart\'{\i}nez--Fern\'andez}
 \affiliation{National Centre for Nuclear Research (NCBJ), Warsaw, Poland}
\author{\'Alvaro Pastor--Guti\'{e}rrez}

\affiliation{Institut f{\"u}r Theoretische Physik,
    Universit{\"a}t Heidelberg, Philosophenweg 16,
    69120 Heidelberg, Germany}

\begin{abstract}

We present a new computation in a field-theoretical model of Coulomb gauge QCD of the first radial and angular excitations of a $qqq$ system  in a SU(3) flavor singlet state, $\Lambda_\text{S}$.  The traditional motivation for the study is that the absence of flavor singlets in the lowest-lying spectrum is a direct consequence of the color degree of freedom.
(The calculation is tested with decuplet baryons $\Delta(1232)$ 
and $\Omega(1672)$.) 
We also analyze decay branching fractions of the flavor singlet baryon for various masses with the simplest effective Lagrangians.

\end{abstract}

\maketitle

\section{Introduction}

\subsection{Color confinement and the three-quark baryon singlet $\Lambda_\text{S}$}\label{subsec:confinementsinglet}

When one examines the empirical basis for ``Confinement'', it is easy to come across studies of ``Quark Confinement'' since fractional charges are a feasible target for searches in Millikan-type experiments~\cite{Perl:2009zz} or at high-energy accelerators~\cite{Bergsma:1984yn}.  However, the theoretical concept that makes sense is rather ``Color Confinement'' (for which the experimental evidence is not so unquestionable~\cite{HidalgoDuque:2011je}, since color leaks by neutral gluons have, surprisingly, not been purposedfully constrained).

Though color is not a useful quantum number for  hadron classification, as we believe they are all color singlets, there are effects due to color in hadron spectroscopy. For example, the predicted Regge trajectories of meson and baryon resonances have different slopes, which can be traced to the $4/3$ versus $2/3$ color factors in gluon exchange between $qq$ and $q\bar{q}$ pairs. Furthermore, one can also think of the  $\pi_0\to\gamma\gamma$ decay, sensible to $N_c$.

This article is driven by our curiosity on the following classic  statement at the root of the quark model and QCD, for which we collect extant evidence from experiment and theoretical computations, including new ones. 
A $qqq$ baryon configuration must be a color singlet, if color is confined; this is achieved by the antisymmetric $\sqrt{\frac{1}{3!}}\epsilon_{ijk} B^\dagger_i B^\dagger_j B^\dagger_k \arrowvert \Omega \rangle $ color wavefunction 
(the quark creation operators  $B^\dagger$ and  color vacuum state $\arrowvert \Omega \rangle$ will be modelled in BCS approximation in section~\ref{sec:singletmass} below).

The antisymmetry of the color wavefunction forces the visible degrees of freedom (spin, orbital angular momentum, quark flavor and eventually radial-like excitations), due to the fermionic nature of spin $\frac{1}{2}$ quarks, to be in a totally symmetric wavefunction; quite unlike nucleon wavefunctions in nuclei or electron wavefunctions in atomic, molecular or solid-state physics.

Chromomagnetic interactions are large in QCD, so one expects (as is typical in hadron physics) that states with lower total angular momentum, $J$, have smaller masses. 
The energy of $qqq$ baryon configurations should be smallest if the spatial degrees of freedom could all be in an s-wave, and also in the lowest radially excited state, that is, 
\begin{equation} \label{spatialwf}
\psi_{\rm spatial} = \prod_{i=1}^3 Y_0^0(\hat{\bf k}_i) R_0(\arrowvert {\bf k}_i \arrowvert) \ , \
\ \sum_{i=1}^3 {\bf k}_i={\bf 0}
.
\end{equation}
Since this is a completely symmetric wavefunction, the remaining product of spin and flavor degrees of freedom must also be in a totally symmetric configuration. 
This means that the lowest two multiplets in the baryon spectrum are Gell-Mann's flavor octet and decuplet~\cite{Close:1979bt,Halzen:1984mc} that combine mixed-symmetry flavor and spin wavefunctions (the octet) and completely symmetric spin and flavor wavefunctions (the decuplet).

The empirical consequence of this quantum wavefunction organization is the absence of a $qqq$ flavor singlet in the lowest-lying spectrum; its antisymmetry would require an antisymmetric spin wavefunction for the spin-flavor product to be symmetric. As it is not possible to antisymmetrize three quarks with only two degrees of freedom (one would be repeated), a flavor singlet with the color degree of freedom requires a spatial-wavefunction excitation (so that part can separately be antisymmetrized). This excitation raises the flavor-singlet mass. Schematically,
\begin{equation}
\psi^A_{\rm qqq} = \psi^A_{\rm color}\otimes \psi^A_{\rm flavor}\otimes 
(\psi_{\rm spin}\otimes \psi_{\rm radial}\otimes \psi_{{\rm orbital}\ L})^A\ ,
\end{equation}
where the $A$ superindex indicates each of the parts that need to separately be antisymmetric. 
There are several ways of achieving antisymmetry of the last parenthesis, and the resulting lowest-energy $qqq$ wavefunctions are explicitly constructed in section~\ref{sec:wfs} below.

It is therefore of theoretical interest to be able to identify a state which coincides, in all or in a good part, with the three-quark antisymmetric-flavor singlet configuration.This is to be found within the $(uds)$ $\Lambda$ hyperon spectrum that contains $\Lambda_\text{S}$, the possible singlets.

To discuss the lightest of those flavor singlet states, in this study we  consider baryons with just one quantum of excitation.

\subsection{Excited $\Lambda$ spectrum} \label{subsec:spectre}

The ground state $\Lambda$ hyperon is well assigned to Gell-Mann's octet: therefore, and as expected, the search for a singlet needs to concentrate on excited states. 
There are two prominent low-energy $\Lambda$ excitations, the $J^\pi=\frac{3}{2}^-\ \Lambda(1520)$ and the $J^\pi=\frac{1}{2}^-\ \Lambda(1405)/\Lambda(1380)$ double system, both with negative parity~\cite{Review:2016}. But as we will show later in section~\ref{sec:wfs}, one expects a $qqq$ singlet configuration with only one quantum of excitation in the $\frac{1}{2}^+$ sector, so we briefly comment on all three channels here.

\subsubsection{$J^\pi=\frac{1}{2}^-$}

The first apparent excitation of the $\Lambda$ is the $S$-wave $\Lambda(1405)$ system, widely believed to be formed by two particles of equal quantum numbers~\cite{Jido:2003cb}  (see, more recently,~\cite{Meissner:2020khl}) mixed from a singlet and two octets with $J^\pi=\frac{1}{2}^-$.
In that classic work, the limit of exact $SU(3)$ symmetry reveals that one of the particle poles, at 1450 MeV, corresponds to a singlet. This pole is generated by the dynamics of the $N-K$
interaction (two octets can yield a singlet irreducible representation of $SU(3)$). Upon breaking $SU(3)$, however, it mixes with the $\Lambda_8$ octets and goes down in mass to 1390 MeV. 

In lattice gauge theory, a state compatible with this $\Lambda(1405)$ was found to give a strong signal with a flavor-singlet interpolating operator~\cite{Engel:2013lea}, making it the lightest solid candidate to belong to the $\Lambda_\text{S}$ singlet family; but how much of the genuine $qqq$ singlet is therein (and how much corresponds to molecular-like configurations, $N\bar{K}$ for example) remained unclear. The answer to this question, as given by~\cite{Hall:2014uca}, is that, at physical pion masses, the state is mostly an antikaon-nucleon bound molecule (as earlier discussed for a long time). Unfortunately, because the interpolator used is an ideally mixed $uds$ configuration, both singlet and octet can contribute to this lattice signal, so the flavor representation or mixing (under scrutiny here) is not extracted. Interestingly, for unphysical pion masses of order the kaon mass or higher, the lattice state becomes an intrinsic (presumably $qqq$) state, but then its mass is in the 1.7-1.8 GeV range, 400 MeV above data.

Next, in one of the Graz quark model computations~\cite{Melde:2008yr}, the Goldstone Boson Exchange (GBE) model (in which quarks exchange pions instead of gluons), the computed mass fits the assignment of $\Lambda(1405)\to \Lambda_\text{S}$, see table~\ref{tab:lattice}, but this model is less widely accepted to represent quark interactions than their One Gluon Exchange (OGE) model that yields a higher mass: this is in agreement with the lattice result of~\cite{Engel:2013lea},  but now too high respect to the experimental datum.

There are two further relatively clear $\frac{1}{2}^-$  excitations at 1670 and 1800 MeV, completing the picture of a singlet and two octets from the meson-nucleon molecule picture and also quark model expectations. It is a fair question to ask how is the $qqq$ flavor singlet distributed among these three states, if at all: the lowest states seem very much influenced by the baryon-meson configuration, and the higher ones have traditionally been assigned to non-singlet multiplets.

\subsubsection{$J^\pi=\frac{3}{2}^-$}

The second well-known $\Lambda$ excitation appears at slightly higher energy above the $KN$ threshold, the $\Lambda(1520)$, which  is a very prominent peak~\cite{Pauli:2019ydi} with $J^\pi=\frac{3}{2}^-$, decaying to both $\Sigma\pi$ and $N\bar{K}$ channels. The lattice computation (typical of what would be a pure $qqq$ state) yields a mass of 1950 MeV in this channel, remarkably higher. The Graz quark models  are closer to the experimental mass.

This is a general pattern: Lattice gauge theory data~\cite{Lin:2011ti,Lin:2008rb}  shows a $\Lambda$ spectrum that is systematically too high respect to the experimental one. A likely reason is that the higher than physical pion mass employed in lattice simulations decouples the meson-nucleon channel, returning the energy of the would-be three-quark core. In this way, our own $qqq$ computation in the NCState Coulomb gauge model, presented below in section~\ref{sec:singletmass}, should more naturally be compared to lattice data than to experimental data. This is shown in table~\ref{tab:lattice}.

There is a second resonance with these quantum numbers, $\Lambda(1690)$, that is usually assigned to a baryon octet~\cite{Guzey:2005rx}.

\subsubsection{$J^\pi=\frac{1}{2}^+$}

If the singlet is searched for with the same $J^\pi$ quantum numbers as the ground-state $\Lambda$, the internal $qqq$ structure needs to be assigned  a radial-like excitation. 

There are two experimentally known resonances, $\Lambda(1600)$ and $\Lambda(1810)$, though this second one apparently is not strictly needed to improve the global fit quality~\cite{Sarantsev:2019xxm}. It is however the one that the Graz group considers the most likely singlet candidate~\cite{Melde:2008yr} in view of their calculations.

The Dyson-Schwinger (DSE) computation~\cite{Qin:2019hgk} predicts a $uds$ excitation with $\frac{1}{2}^+$ around $1475$ MeV, though the authors believe that model dependence is dragging it downwards: if they opt for artificially weakening their kernel interaction, by less than 10\%, they bring it up to 1580 MeV, in line with other $qqq$  approaches. This is marked with an asterisk in table~\ref{tab:lattice}.

\begin{table*}
\caption{
The lattice QCD data from~\cite{Lin:2011ti,Lin:2008rb} shows a spectrum of $\Lambda$ resonances at substantially larger mass than the experimental states. This is natural taking into account that the pion mass $m_\pi$ is taken in the interval 300-700 MeV in the lattice simulations  (in effect closing the decay phase space), corrected by a linear extrapolation $M_\Lambda \propto a + b m_\pi^2$ to the physical 138 MeV mass. Thus, our Coulomb-QCD model computation of pure $qqq$ states (selected in a flavor singlet configuration) is more comparable to this lattice calculation than directly to the experimental spectrum. Our restriction of the flavor to a singlet is likely raising the mass, as can be seen comparing to other theoretical approaches. All masses rounded off to 5 MeV. \label{tab:lattice}} 
\begin{tabular}{|c|c|c|c|c|c|}\hline
Experimental & \multicolumn{4}{c|}{Mixed $uds$ configurations} & Singlet configuration \\ \hline
 $\Lambda$ candidates & Lattice & Graz & Bonn & Dyson- &  Coulomb gauge \\
 &         & models  & model~\cite{Ronniger:2011td} (\cite{Loring:2001kx}) & Schwinger~\cite{Qin:2019hgk}  & model (this work) \\ \hline
$\Lambda(1380+1405)\frac{1}{2}^-$ & 1600~\cite{Engel:2013lea}  & 1555 (GBE) & 1620 (1511) & 1315 &  \\
$\Lambda(1670)\frac{1}{2}^-$ & 1450~\cite{Hall:2014uca}   & 1630 (OGE) & 1695(1635) & ($1580^*$) & \bf $1800\pm 200$   \\
$\Lambda(1800)\frac{1}{2}^-$ &                            &     & 1830(1774) &       & \\ \hline
$\Lambda(1520)\frac{3}{2}^-$ & 1950  &  1555 (GBE) & 1595 (1500) & & \bf$1700\pm 200$ \\
$\Lambda(1690)\frac{3}{2}^-$ &       &  1630 (OGE) & 1710 (1650) & & \\ \hline
$\Lambda(1600)\frac{1}{2}^+$ & 1900~\cite{Nakajima:2001js}    &  1625 (GBE) &1590(1665)  & 1475 & \bf $2400\pm 150$ \\
$\Lambda(1810)\frac{1}{2}^+$ &       &  1745 (OGE) & 1790(1750) & ($1580^*$) & \\  \hline
\end{tabular}
\end{table*}

Several other aspects of the table merit comment. We quote two different instanton-interacting Bonn quark-model computations from~\cite{Ronniger:2011td} and~\cite{Loring:2001kx}. They differ in that the later employs a flavor-independent kernel, whereas the former, a later computation, introduces a flavor dependence to improve agreement with the data. This is achieved, but then disagreement with lattice data (that should better represent the $qqq$ configuration) arises.

\subsection{Flavor structure}

As $SU(3)$ symmetry is not exact, octet-singlet flavor mixing (and eventually, even with higher representations) is expected to happen. Of mesons  we know, for example, that the $\omega$ is purely $u\bar{u}+d\bar{d}$ while the $\phi$ is almost entirely $s\bar{s}$ (ideal mixing), while the pseudoscalar $\eta,\ \eta'$ pair is in a differently mixed configuration, though not purely octet-singlet; ground state baryons are however widely believed to be in a rather good octet configuration.
Remarkably, the Gell-Mann-Okubo formulae for the octet $\frac{1}{2}^+$ baryons are accurate~\cite{Guzey:2005rx} to $O(15{\rm MeV})\sim 1-2\%$ in spite of the possible mixing. The mixing seems to be small, and because its dependence in the controlling $\sin \theta_{1-8}$ is quadratic, the angle is difficult to extract with precision.

Turning to the excited states with which we here deal, assigning the $\frac{3}{2}^-$ $\Lambda(1520)$ to be a pure flavor-singlet baryon is problematic because of its decay to $\Sigma(1385)\pi$, as ${\bf 1}\not\to {\bf 10}+ {\bf 8}$~\cite{Guzey:2005rx}, so that invoking mixing with a higher resonance of equal spin-parity, presumably the 1690, belonging to a flavor octet according to other work ~\cite{Melde:2008yr} seems necessary.

Also in the negative parity sector~\cite{Nakajima:2001js},
an interesting quenched lattice calculation that separately analyzed the correlators,
found very similar octet and singlet masses for the $\Lambda \frac{1}{2}^-$, (and this around 1.6 GeV in agreement with~\cite{Lin:2011ti}). That could indicate that in that channel an octet and a singlet should appear almost degenerate and  mixed, which seems to be the case for the $\Lambda(1380)-\Lambda(1405)$ system (though at a smaller mass consistent with a strong nucleon-meson open channel influence). However, the extent to which this system can be considered $qqq$ remains questionable: this system might be mixed, but not be so relevant for our thrust.

There does not seem to be much information in the octet-singlet comparison for higher excitations nor for the $\frac{3}{2}^+$ channel, but we can draw from the active field on $\Lambda_\text{c}$ and $\Lambda_\text{b}$ spectroscopy: for example, an excited likely $\Lambda_\text{b} \frac{1}{2}^+$ candidate has just been reported~\cite{Azizi:2020ljx} (see for example~\cite{Ebert:2011kk} for quark-based theory discussion thereof).

\begin{figure*}
\includegraphics[width=\textwidth]{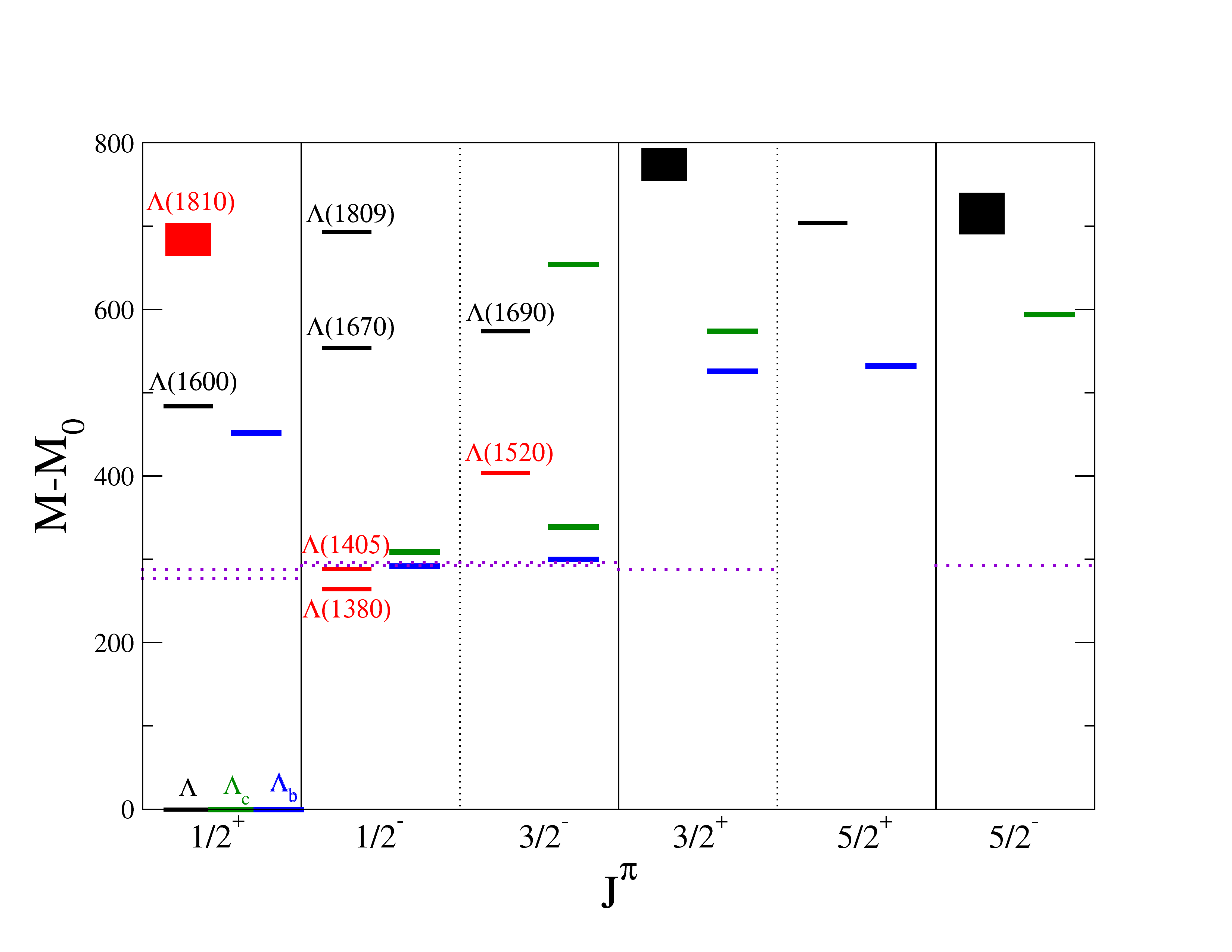}
\caption{Comparative of the $\Lambda$ spectrum with the known $\Lambda_\text{c}$ and $\Lambda_\text{b}$ states, that mark ideal $udc$ and $udb$ valence mixing, without $SU(3)$ symmetry. To discuss the congruence of the spectra, we plot $M-M_0$ with $M_0=1116,\ 2286,\ 5620$ MeV, for s,c,b respectively, (and in the same spirit the second shell levels of Helium with $A=3$ are also marked as dotted lines, with energies rescaled to match the $\Lambda_\text{b}\frac{1}{2}^-$). The $\Lambda_\text{S}$ singlet candidates highlighted by the Graz group~\cite{Melde:2008yr} (1810, 1405/1380, 1520) are displayed in red online. \label{fig:spectrum}}
\end{figure*}

In figure~\ref{fig:spectrum} we have displayed the $\Lambda$ spectrum against the $\Lambda_\text{c}$, $\Lambda_\text{b}$ (and marked the rescaled second shell of the $^3He$ atomic $A-e-e$ three-body system) as a benchmark.

The ground-state energy of all of them has been subtracted, so that only the excitation energy is seen in the plot.

As has been known for long, the charmed/charmonium and bottomed/bottomonium system have congruent spectra on this type of Grotrian diagrams (see {\it e.g.} \cite{Rosner:2006jz}). This is due to the shape of the interquark Cornell linear+Coulomb potential that looks, in the momentum range where both pieces are of comparable magnitude, somewhat like a logarithmic potential (that would show actual matching of the spectra upon subtracting the ground state).

The heavy-baryon spectrum shown in the figure should correspond to the pure valence or ideally-mixed configurations $(ud)c$, $(ud)b$ with little or no further flavor configuration mixing expected to affect the heavy quark, which is distinguishable and more localized than the others due to its large mass.
The figure teaches us that the splittings to the ground state are generically larger for the strange $\Lambda$ states than their heavy-quark counterparts, probably due to these being less relativistic, but they seem rather comparable. This suggests perhaps that one quantum of excitation costs a similar amount whether concentrated in a part of the system such as in $(ud)b$ or distributed through the three quarks such as in $(uds)_8$ or $(uds)_1$. Our findings within the Coulomb gauge approach (see again table~\ref{tab:lattice}) would however indicate that the singlet $qqq$ configuration can be a bit heavier in the Cornell linear+Coulomb potential. 

Flavor mixing in the QCD Hamiltonian resides exclusively in the quark mass matrix, that for exact isospin symmetry can be written as
\begin{equation}
\label{matrizmasa}
[M] = \frac{m_u+m_d+m_s}{3} \mathbb{I} - \frac{m_s-(m_u+m_d)/2}{\sqrt{3}} \lambda_8\ .
\end{equation}
Since the second term, not respecting $SU(3)$ symmetry, is in the octet representation as Gell-Mann's $\lambda_8$ matrix reveals, singlet and octet hyperon representations can be mixed, but not singlet and decuplet ones. The Bonn group has extended\cite{Ronniger:2011td} their earlier work to explore additional sources of flavor violation in an effective $qqq$ Hamiltonian that is meant to incorporate effects of meson exchange among quarks in a potential. Because mesons have rather different masses, this potential is strongly flavor dependent. 

In our own calculation in section~\ref{sec:singletmass} we have kept the canonical interaction with the global symmetries of QCD, so that our flavor-violation is reduced to the quark mass matrix in Eq.~(\ref{matrizmasa}). Moreover, because dynamical chiral symmetry breaking means that the running quark mass decreases with the scale (unlike in constituent models where the mass is fixed; this point will be clear in figure~\ref{fig:gapeq} below), our approach should have significantly less flavor mixing than others. This probably oversimplifies a more complex physical picture, but makes discussion of the flavor singlet configuration, whose absence in the low spectrum is the telltale of color, more straightforward.

To conclude this section, though it is often manifested that quark models cannot be used for precision work, which is fair criticism given the uncontrolled approximations that are needed to reduce them to manageable calculations, the prediction of the quantum numbers for the lowest $\Lambda$ excitations is spot on: indeed, the first excitation in the quark model can have quantum numbers $J^\pi = \frac{1}{2}^-, \frac{3}{2}^-, \frac{1}{2}^+$ 
as demonstrated next in section~\ref{sec:wfs}. These happen to be the quantum numbers of the first few experimentally detected states, a refreshing agreement, so there are several possible candidates to $\Lambda_\text{S}$ singlet.


\section{Construction of variational wavefunctions for lowest-lying $qqq$ flavor-singlet baryons} \label{sec:wfs}

As discussed in subsection~\ref{subsec:confinementsinglet},
the $SU(3)$ flavour-singlet $qqq$ baryon is in an antisymmetric flavor combination, and because of the confinement hypothesis, also antisymmetric in color, with Fermi statistics necessarily leaving an antisymmetric spin and spatial wavefunction of its valence quarks. 

We employ configurations with well defined total angular momentum $J:=|{\bf J}|$, third component $\langle J_z \rangle$ and parity $\pi$. To obtain $J$ we combine the doublet representations from spin and spatial quantum numbers, which are a set of mixed-symmetric and mixed-antisymmetric configurations. 
The spatial part carries standard spherical harmonics $Y_l^m(\hat{k}_i)$  with $\hat{\bf k}_i = {\bf k}_i/k_i$, ${\bf k}_i$ being the quark momentum and $k_i = |{\bf k}_i|$ its modulus.

With only one quark excited above the ground state we can construct three different $J^\pi$ combinations with totally antisymmetric spin-spatial states,
\begin{enumerate}
\item $1/2^+$ with a radial-like excitation $R_1(k)Y_0^0(\hat{k})$ (see Eq.~(\ref{1/2+}) below).
\item $3/2^-$ with an angular excitation $R_0(k)Y_1^m(\hat{k})$  (see Eq.~(\ref{3/2-}) below).
\item $1/2^-$, also with an angular excitation $R_0(k)Y_1^m(\hat{k})$   (see Eq.~(\ref{1/2-}) below).
\end{enumerate}

The two quarks that remain in the ground state  are naturally assigned wavefunctions  $R_0(k)Y_0^0(\hat{k})$. The three quarks are then antisymmetrized without concern to their mass/flavor (the flavor wavefunction is by construction antisymmetric itself). It is this step that suppresses any singlet-octet flavor mixing that may lower the mass: all results in this work refer to the pure flavor-singlet configuration.
In a $\Lambda_\text{c}$ or $\Lambda_\text{b}$ baryon, one quark is in a definite flavor state; not here, all three have some probability amplitude of being the strange quark.

As explained in subsection~\ref{subsec:confinementsinglet}, we need to introduce that spatial excitation in order to build an antisymmetric spin-spatial wavefunction appropriate for the singlet baryon. Hence, we have two spin quantum states ($\pm 1/2$), and either two radial states  ($n_r \in \{0, 1\}$ ground/excited) or two angular states ($l \in \{0, 1\}$). 
In each of these spaces we have therefore a doublet of an $SU(2)$-like group and for three quarks we have a tensor product, which furnishes a reducible representation 
thereof,
\be
{\bf 2} \otimes {\bf 2} \otimes {\bf 2} = {\bf 4} \oplus {\bf 2} \oplus {\bf 2}\ .
\ee

The quadruplet ${\bf 4}$ in the resulting direct-sum decomposition is totally symmetric under permutations of the three quarks, so joint antisymmetry of the spin-space wavefunction demands the usage of the two doublets ${\bf 2}$. These can be related to a couple of mixed-symmetric ({\it MS}) and mixed-antisymmetric ({\it MA}) states for spin $\chi_{MS, MA}(S, M_S)$
\begin{eqnarray}
\label{sms1}\chi_{MS}\left(\frac{1}{2}, \frac{1}{2}\right) & = & \frac{1}{\sqrt{6}}\left(2 |\uparrow\uparrow\downarrow \rangle - |\uparrow\downarrow\uparrow\rangle - |\downarrow\uparrow\uparrow\rangle\right)\nonumber \\
\chi_{MS}\left(\frac{1}{2}, -\frac{1}{2}\right) & = & \frac{1}{\sqrt{6}}\left(2 |\downarrow\downarrow\uparrow\rangle - |\downarrow\uparrow\downarrow\rangle - |\uparrow\downarrow\downarrow\rangle\right)\nonumber \\
\chi_{MA}\left(\frac{1}{2}, \frac{1}{2}\right) & = & \frac{1}{\sqrt{2}}\left(|\uparrow\downarrow\uparrow\rangle - |\downarrow\uparrow\uparrow\rangle\right)\nonumber \\
\label{sma2}\chi_{MA}\left(\frac{1}{2}, -\frac{1}{2}\right) & = & \frac{1}{\sqrt{2}}\left(|\downarrow\uparrow\downarrow\rangle - |\uparrow\downarrow\downarrow\rangle\right)
\end{eqnarray}
\\
a construction that is immediately exported to orbital angular states $|L, M_L\rangle_{MS, MA}$
\begin{eqnarray}
\label{ams}|1, M_L\rangle_{MS} & = & \frac{1}{\sqrt{6}}(2 |001_{M_L}\rangle - |01_{M_L}0\rangle - |1_{M_L}00\rangle)\nonumber \\
\label{ama}|1, M_L\rangle_{MA} & = & \frac{1}{\sqrt{2}}(|01_{M_L}0\rangle - |1_{M_L}00\rangle)
\end{eqnarray}
\\
and to orbital radial ones $|{\rm rad} \rangle_{MS, MA}$
\begin{eqnarray}
\label{rms}|{\rm rad}\rangle_{MS} & = & \frac{1}{\sqrt{6}}(2|001\rangle - |010\rangle - |100\rangle)\nonumber \\
\label{rma}|{\rm rad}\rangle_{MA} & = & \frac{1}{\sqrt{2}}(|010\rangle - |100\rangle) \ .
\end{eqnarray}
\\
Combining these states, we can form antisymmetric combinations of the spin-orbital ones
\begin{multline}
|M_S; M_L\rangle = \frac{1}{\sqrt{2}}(\chi_{MS}(1/2, M_S) |1, M_L\rangle_{MA} \\
- \chi_{MA}(1/2, M_S) |1, M_L\rangle_{MS})
\end{multline}
or of the spin-radial ones
\begin{multline}
    |M_S;{\rm rad}\rangle = \frac{1}{\sqrt{2}}(\chi_{MS}(1/2, M_S) |{\rm rad}\rangle_{MA}  \\
    - \chi_{MA}(1/2, M_S) |{\rm rad}\rangle_{MS})
\end{multline}

Once fully antisymmetric representations of the quark permutation group are achieved, the Clebsch-Gordan coefficients assist in obtaining antisymmetric spin-spatial wavefunctions with well-defined $J^\pi$. As advanced, with only one quantum excitation there are three cases, in agreement with earlier work~\cite{Melde:2008yr}
\be\label{1/2+}
    |1/2^+\rangle = |M_S = 1/2; {\rm rad}\rangle
\ee
\be\label{3/2-}
|3/2^-\rangle = |M_S = 1/2; M_L = 1\rangle
\ee
\begin{eqnarray}
|1/2^-\rangle & = & \sqrt\frac{2}{3}|M_S = -1/2; M_L = 1\rangle \nonumber \\
& & + \frac{1}{\sqrt{3}}|M_S = 1/2; M_L = 0\rangle\ . \label{1/2-}
\end{eqnarray}

This resulting collection of quantum numbers is used to prepare the necessary effective Lagrangians to study branching fractions of flavor-singlet baryon decay at the hadron level in section~\ref{decays}  below.

Additionally, these wavefunctions are also injected into the Rayleigh-Ritz variational principle $\langle \psi^\dagger_S \arrowvert H \arrowvert \psi_S \rangle \leq E_S \langle \psi^\dagger_S \arrowvert  \psi_S \rangle $ with the quark-level Hamiltonian specified below in section~\ref{sec:singletmass}.

Symmetry considerations do not fix the wavefunctions entirely, leaving what are usually  called ``radial'' excitations in Eq.~(\ref{rms}) (concept that makes sense once the 3-body variables have been fixed).

We have employed three different variational radial Ans\"atze in closed analytical form. Each is a family of functions with up to three variational parameters $\rho_i$, one  for each quark.
Because we impose the center of mass $\sum{\bf k}_i={\bf 0}$ condition, one of the parameters is redundant; we prefer to dedicate the additional computer time spent in the redundancy than further complicating the wavefunctions. They read
\be
R_n^{(0)}(k) = \begin{cases}
    \left[\left(\frac{k}{\rho}\right)^2 + 1 \right]^{-2},\ \mbox{ for $n$ = 0}\\
    \frac{2}{\left[\left(\frac{k}{\rho}\right)^2 + 1\right]^{2}} + \frac{\left(\frac{k}{\rho}\right)^2 -\frac{3}{4}}{\left[\left(\frac{k}{\rho}\right)^2 + \frac{1}{4}\right]^3},\ \mbox{for $n$ = 1} \label{radial0}
    \end{cases}
\ee
\be
R_n^{(1)}(k) = \begin{cases}
    \frac{k}{\rho}e^{-\left(\frac{k}{\sqrt{2}\rho}\right)^2},\ \mbox{for $n$ = 0}\\
    \left[ 2\left(\frac{k}{\rho}\right)^3 - 3\frac{k}{\rho}     \right]e^{-\left(\frac{k}{\sqrt{2}\rho}\right)^2},\ \mbox{for $n$ = 1}
    \end{cases}
\ee
\be
R_n^{(2)}(k) = \begin{cases}
    \left[\frac{3}{2} - \left(\frac{k}{\rho}\right)^{2}  \right]e^{-\left(\frac{k}{\sqrt{2}\rho}\right)^2},\ \mbox{for $n$ = 0}\\
    \left[\frac{15}{4} - 5\left(\frac{k}{\rho}\right)^{2} + \left(\frac{k}{\rho}\right)^{4}  \right]e^{-\left(\frac{k}{\sqrt{2}\rho}\right)^2},\ \mbox{for $n$ = 1}
    \end{cases}
\ee

The first one corresponds to a hydrogen-like wavefunction; the second to a one-dimensional harmonic oscillator; and the third one is related to the three-dimensional harmonic oscillator. These wavefunctions can be employed to apply the variational principle to any appropriate QCD or QCD-like Hamiltonian.

Finally, we employ a fourth wavefunction family which is implemented as a numeric table to be interpolated. The table is obtained by solving the $1^{--}$ meson problem with the same Hamiltonian and potential parameters later used for the three-body problem. 
\be
R_n^{(3)}(k) = \begin{cases}
\psi_{\rho}(k),\ \mbox{for $n$ = 0}\\
\psi_{\rho'}(k),\ \mbox{for $n$ = 1}
\end{cases}
\ee
where the $\rho$ two-body problem was simplified by ignoring $d$-wave or back-propagating Salpeter (Random Phase Approximation) contributions, so that 
the two wavefunctions with $n=0,\ 1$ are adequately orthogonal in the radial $k$ variable (without being concern about the precision reached in the $\rho$ spectrum, that does require the additional contributions). 

The hope is that this wavefunction, by having the correct tails for the interaction, will be able to relax $\langle H \rangle$ somewhat more than the others (this will be shown to be the case in one instance, the $\frac{1}{2}^+$ $\Lambda_\text{S}$, whereas for the other quantum number combinations, $R_n^{(0)}$ performs equally well, so we will quote results therefrom as it is more straighforward).

 But before deploying a specific calculation, we dedicate a section to exploiting our gained knowledge of the possible quantum numbers in the low spectrum to discuss $\Lambda_\text{S}$ decays in the next section~\ref{decays}.

\section{Decay branching ratios of the $SU(3)$--flavor singlet baryon as function of its mass}\label{decays}

In this section we will employ the simplest methods of Effective Theory to learn about the relevant two-- (subsections~\ref{subsec:decayY} and~\ref{subsec:decay2}) and, only for one case, three--body (subsection~\ref{subsec:decay3}) decay widths of the $\Lambda_\text{S}$ baryon, 
without attempting to probe its internal structure, but exploiting flavor symmetry and phase space to relate different decay channels. The overall decay constant of the Effective Lagrangians below, such as Eq.~(\ref{decay2}), (\ref{decay2_3halves}) and (\ref{Lag:3}), will be left undetermined, so 
that predictivity extends only to branching fractions
$\Gamma_i/\Gamma_{\rm total}$. 

This is of interest, from a purely experimental point of view, to eventually understand how well do the existing physical baryons with the same quantum numbers match a pure singlet configuration; but also to explore the influence of the open channels to the seed $qqq$ baryons that quark--approaches produce. Naturality suggests that the imaginary part of the baryon propagator (thus, the decay width) is of similar size to the correction to the real part, shifting its mass (that, at order zero, is seeded by the pure quark calculations discussed below in section~\ref{sec:numeric}). 
This section employs standard notation  of hadron Effective Theory: $B$ will represent the ground--state baryon flavor--octet of Gell--Mann, and $\Phi$ the pseudoscalar meson octet.

\subsection{Two-body decays with a contact Yukawa Lagrangian}\label{subsec:decayY}
We consider first two--body $B\phi$ baryon--meson decays of the singlet $\Lambda_\text{S}$, that is, $\Lambda_\text{S} \rightarrow B +\phi$ decay.

In this subsection we adopt the simplest contact Yukawa Lagrangian density. This is analogous to the analysis of flavor in baryon decays  carried out by Guzey and Polyakov~\cite{Guzey:2005rx} (though they cover the entire spectrum) to which we refer for extensive discussion.
They find the ratios among coupling constants, exclusively based on the flavor structure, given by
$g_{\Lambda_\text{S} NK}:g_{\Lambda_\text{S} \Sigma\pi}: g_{\Lambda_\text{S} \Lambda\eta }:g_{\Lambda_\text{S} \Xi K} = \frac{1}{2}:\sqrt{\frac{3}{2}}:-\frac{1}{2\sqrt{2}}: -\frac{1}{2}$
and whose squares give a first idea of the relative importance of the different two-body decay channels.

Such coefficients are hidden from direct experimental acces due to two problems that are adding up for states of low and moderate mass. The first is the phase-space integral: if states are not too far from the respective thresholds (or even below, with zero width!) the $SU(3)$ relations of the couplings are completely wiped out by the large $SU(3)$-breaking induced by the very different momenta, in turn coming from the decay-product masses by K\"allen's formula $|\textit{\textbf{p}}| = \dfrac{1}{2m_{\Lambda_\text{S}}} \lambda^{1/2} (m_{\Lambda_\text{S}}^2, m_1^2, m_2^2)$. 
This first issue is easily addressed by proceeding to the total width that can be given in numeric form to compare with experiment, 
\begin{equation}
\Gamma (\Lambda_\text{S} \rightarrow 1+2) = \frac{|\textit{\textbf{p}}|}{32 \pi^2 m_{\Lambda_\text{S}}^2} \int | {\mathcal M}|^2 \text{ d}\Omega 
\end{equation}

\begin{figure}
    \includegraphics[width=0.5\textwidth]{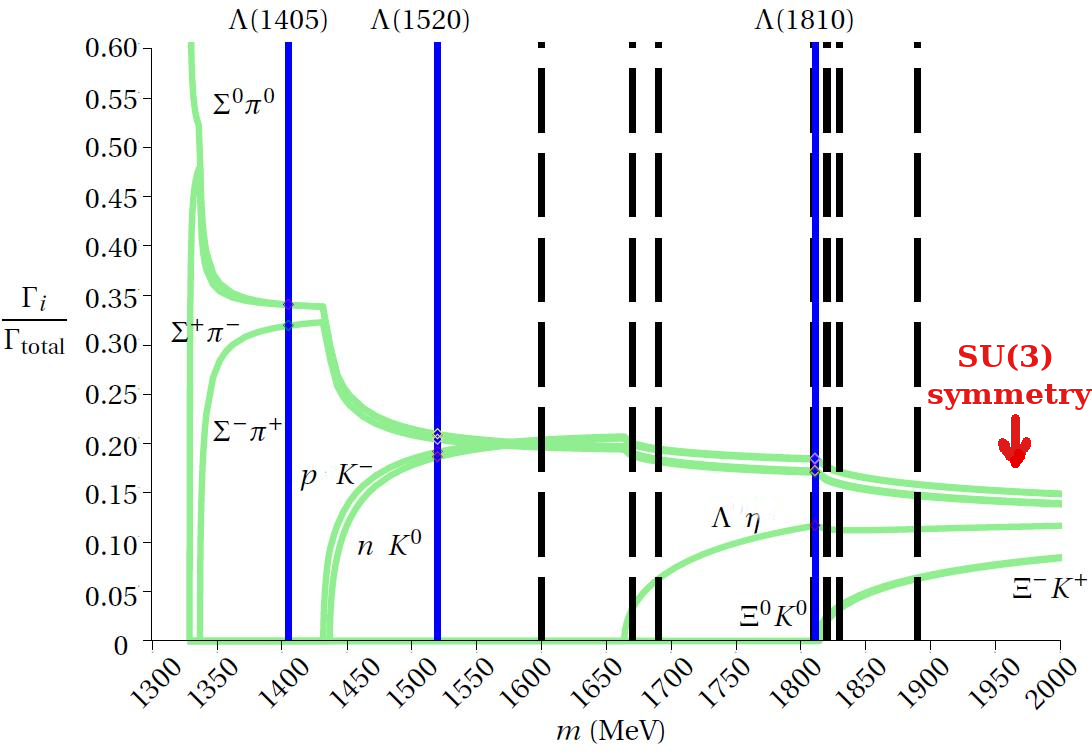}
    \includegraphics[width=0.5\textwidth]{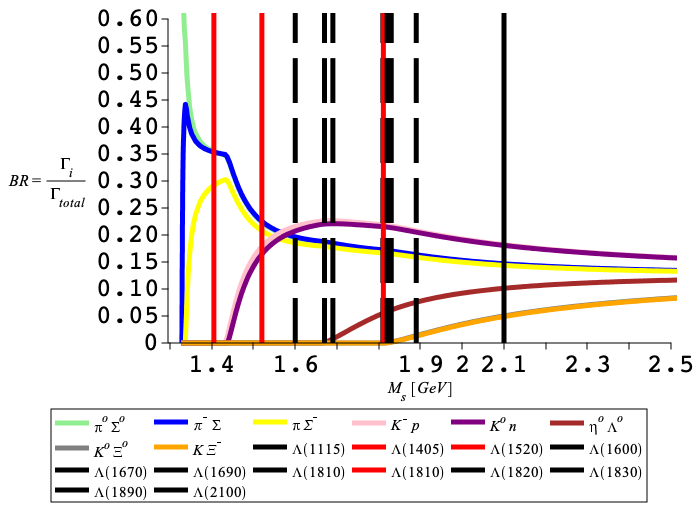}
    \caption{\label{fig:decays2chiral} Branching ratios of the two--body flavor--preserving decay channels of the SU(3) singlet $\Lambda_\text{S}$  as function of $M_s:=m_{\Lambda_\text{S}}$ for $J^\pi = \frac{1}{2}^+$. The top plot shows the case of a constant Yukawa vertex, whereas the bottom plot employs the derivatively coupled amplitude of Eq.~(\ref{decay2_final}). }
    \end{figure}

To get rid of the model-dependence of the $g_i$, we plot $\frac{\Gamma_i}{\sum_{2\rm \ body} \Gamma_i}$ in figure~\ref{fig:decays2chiral} (top plot). The detailed discussion of such plots is postponed to subsection~\ref{subsec:decay2}, but let us note here how all channels tend to a constant (energy-independent) decay fraction at large decaying-particle mass (flavor symmetry) whereas, at low momenta, different phase space makes the various lines immensely different.

The second problem with a constant coupling is that the pion and, to a lesser extent, the Kaon and the eta are quasi-Goldstone bosons, and the construction of chirally symmetric Lagrangians demands that they are derivatively coupled. Constant, momentum-independent couplings, 
can of course be present too, since chiral symmetry is not exact, but the large derivatively-coupled contribution can enhance the apparent $SU(3)$-symmetry breaking of the decay by the same mechanism, the different momenta induced by the different masses.

\subsection{Two-body decays with a derivatively coupled meson}\label{subsec:decay2}
After the brief example of a  constant Yukawa coupling, we proceed to examine the derivatively coupled meson Lagrangian for all three $J^\pi$ combinations of interest for a $qqq$ $\Lambda_\text{S}$.

\paragraph{State with $J^\pi=\frac{1}{2}^+$}
 We use the simplest perturbative Lagrangian with a pion derivative coupling as suggested by the chiral theory of the strong interactions~\cite{Thomas:2001} and $SU(3)$ symmetry,
    \begin{equation}\label{decay2}
        L = -\dfrac{g}{2f_\pi} \overline{\Psi}_{\Lambda_\text{S}} \gamma_\mu \gamma_5 {\rm tr}(\Psi_B \partial^\mu \Phi)
    \end{equation}
    where $g$ is the decay coupling and $f_\pi$ the weak meson decay constants, both flavor independent;
$\Psi_B$ is the octet and  $\Psi_{\Lambda_\text{S}}$ the singlet baryon fields;  and $\Phi$ is the meson field (the flavor trace is taken over the product of the two octet matrices). It yields a matrix element

\begin{eqnarray} 
 \overline{|\mathcal{M}|^2} = \nonumber \\ {\rm Tr}\left(
 (\not \! p_{\Lambda_\text{S}} - \not \! p_{B})\gamma_5
 (\not \! p_{\Lambda_\text{S}} + m_{\Lambda_\text{S}}) \gamma_5 (\not \! p_{ B}-\not \! p_{\Lambda_\text{S}}  ) (\not \! p_{ B} + m_{ B})
 \right) \nonumber \\
\end{eqnarray}

that is especially simple if evaluated in the rest frame of the decaying $\Lambda_\text{S}$ baryon,
    \begin{equation} \label{decay2_final}
        \overline{|\mathcal{M}|^2}= (m_{\Lambda_\text{S}}+m_{B})^2 \bigl((m_{\Lambda_\text{S}}-m_B)^2-m_\phi^2\bigr)
    \end{equation}
    in terms of the respective masses.
    
    The two--body flavor--preserving decay channels of the SU(3) singlet $\Lambda_\text{S}$ are $\Sigma^0 \pi^0$, $\Sigma^+\pi^-$, $\Sigma^- \pi^+$, $pK^-$, $n K^0$, $\Lambda^0 \eta/\eta'$, $\Xi^0 K^0$ and $\Xi^- K^+$.
    Their branching ratios are presented in lower plot of figure~\ref{fig:decays2chiral} (bottom plot).

The vertical solid lines (red online) correspond to the three $\Lambda$ resonances in the 1.4-2 GeV region that are candidates to be (or to contain a sizeable amount of the wavefunction of) the lightest flavor singlet as per the Graz proposed assignment~\cite{Melde:2008yr}. The rest of the lines represent various other $\Lambda_\text{S}$ resonances.
If we take the current $J^\pi$ assignments of the Particle Data Group at face value, $\Lambda(1600)$ and $\Lambda(1810)$ are the lightest relevant ones (subsection~\ref{subsec:spectre} )

To exemplify the use of such graphs, let us focuse on the $\Lambda(1600)$ that the Graz group classified as belonging to a first excited octet with $\frac{1}{2}^+$ including the $N(1440)$ Roper resonance. It corresponds to the first vertical dashed line. From the graph we see that there are five channels with an approximately equal branching fraction, the three $\pi \Sigma$ charge combinations and the two $KN$ ones. This entails a prediction
$\Gamma_{\pi\Sigma}/\Gamma_{KN}\simeq 1.5$ that would be informative in possession of more accurate experimental data (currently, the PDG average is consistent with a broad interval $\Gamma_{\pi\Sigma}/\Gamma_{KN}\simeq 1.1-3.1$).

\paragraph{State with $J^\pi=\frac{1}{2}^-$}

\begin{figure}[h!]
    \includegraphics[width=0.5\textwidth]{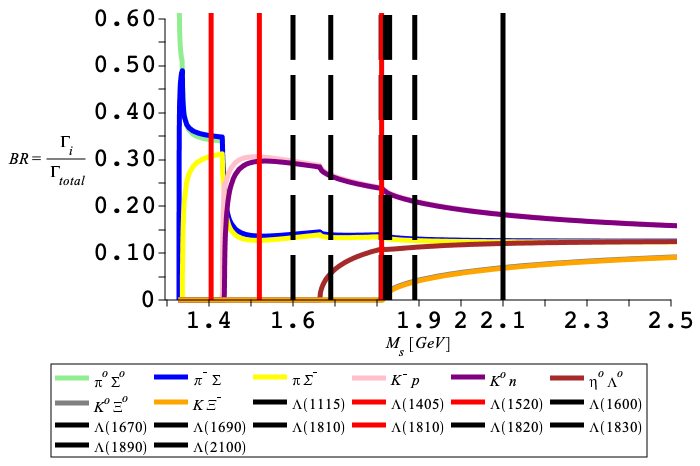}
    \caption{Branching ratios of the two--body flavor--preserving decay channels of the SU(3) singlet $\Lambda_\text{S}$ \label{fig:decays3chiral} as function of $M_s:=m_{\Lambda_\text{S}}$ for $J^\pi = \frac{1}{2}^-$.}
    \end{figure}

If the $\Lambda_\text{S}$ state has parity opposite to the nucleon, the decay vertex equivalent to Eq.~(\ref{decay2}) will lack the $\gamma_5$ so that the total Lagrangian density is parity-even as correspond to the strong force. In that case, Eq.~(\ref{decay2_final}) turns into
    \begin{equation} \label{decay2minus_final}
        \overline{|\mathcal{M}|^2}=  (m_{\Lambda_\text{S}}-m_{B})^2 \bigl( (m_{\Lambda_\text{S}}+m_B)^2 -m_\phi^2 \bigr)\ .
    \end{equation}
The resulting relative 2-body decay intensities are shown in figure~\ref{fig:decays3chiral}.

For example, a singlet state with the mass of $\Lambda(1670)$ would decay in the ratios
$KN:\pi\Sigma:\eta\Lambda=1:0.7:0.1$ approximately, whereas the experimental quotients seem to be  $1:0.3-0.7:0.4-1.4$. Thus, current experimental data is not yet at the precision level where it could exclude this particle from a singlet assignment just from its decays (one needs to resort to Gell-Mann-Okubo type arguments seeing whether the state fits well inside a complete baryon octet or not).

\paragraph{State with $J^\pi=\frac{3}{2}^-$}

The third basic excitation that can form a $qqq$ flavor singlet has spin $\frac{3}{2}$. This requires the use of a higher representation of the Lorentz group than conventional spinors: 
a convenient formalism is that of Rarita and Schwinger. While it is usually not covered in basic treatments, it is somewhat widely known, so we compromise by giving the detail of the calculation but relegating it to appendix~\ref{ap:RS}. The decay vertex is now
 \begin{equation}\label{decay2_3halves}
        L = g_{\Lambda_\text{S} B \phi}{\rm tr} \left[ \overline{\Psi}_B\left( P^{3/2}\right)^{\mu\nu} 
        \left(\Psi_{\Lambda_\text{S}\ RS}\right)_\nu  \partial_\mu \Phi \right] +{\rm h.c.}
    \end{equation}
where $\left(\Psi_{\Lambda_\text{S}\ RS}\right)_\nu$ is the Rarita-Schwinger collection of spinors described in appendix~\ref{ap:RS} and 
$\left( P^{3/2}\right)^{\mu\nu}$ the projector necessary to pick up the spin $\frac{3}{2}$ component therefrom. The flavor trace has also been taken so the Lagrangian density is a flavor singlet.
This leads to a squared matrix element 
\begin{eqnarray}\label{RStrace}
 \overline{|\mathcal{M}|^2} = 
  (p_{\Lambda_\text{S}} - p_{B})_\mu
   (-p_{\Lambda_\text{S}} + p_{ B})_\nu
 \nonumber \\ {\rm Tr}\left(
\left( P^{3/2}\right)^{\mu\nu}
  (\not \! p_{\Lambda_\text{S}} + m_{\Lambda_\text{S}})
  (\not \! p_{ B} + M_{ B})
 \right) 
 \nonumber \\ \ .
\end{eqnarray}
Since the projector 
$P^{3/2}$ somewhat complicates the calculation, we have carried it out with the help of the symbolic manipulation system FORM~\cite{Ruijl:2017dtg}. We organize the result as a power-series expansion in $m_{\Lambda_\text{S}}$, yielding
\begin{widetext}
\begin{eqnarray}\label{decay3hminus_final}
\overline{|\mathcal{M}|^2} &\propto &
\frac{1}{3} m^4_{\Lambda_\text{S}} +\frac{2}{3} m^3_{\Lambda_\text{S}} m_B - m^2_{\Lambda_\text{S}} \left( m_\phi^2+
\frac{1}{3}m_B^2\right) \nonumber \\
& & -\frac{4}{3} m_{\Lambda_\text{S}} \left( m_\phi^2+
m_B^2\right)m_B - \left( \frac{1}{3}m_B^4+\frac{2}{3} m_B^2 m_\phi^2-m_\phi^4\right) \nonumber \\
& & + \frac{2}{3} \frac{m_B}{m_{\Lambda_\text{S}}} \left(m_B^2-m_\phi^2 \right)^2 - \frac{1}{m_{\Lambda_\text{S}}^2}
\left( \frac{1}{3}m_\phi^6 -m_B^2m_\phi^4 +m_B^4m_\phi^2
-\frac{1}{3} m_B^6 \right)
\end{eqnarray}
\end{widetext}

\begin{figure}[h!]
    \includegraphics[width=0.5\textwidth]{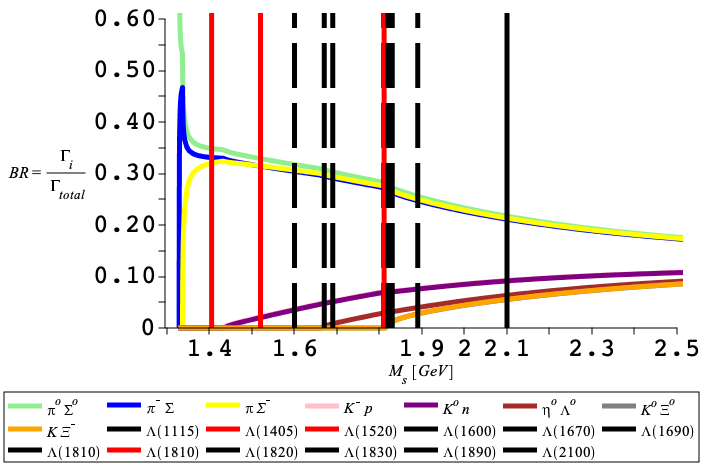}
    \caption{Branching ratios of the two--body flavor--preserving decay channels of the SU(3) singlet $\Lambda_\text{S}$ \label{fig:decays32} as function of $M_s:=m_{\Lambda_\text{S}}$ for $J^\pi = \frac{3}{2}^-$.}
    \end{figure}

After folding it with phase space, the resulting relative two-body branching fractions are plotted in figure~\ref{fig:decays32}, with conventions equal to those of figure~\ref{fig:decays2chiral}.
    
We find remarkable that the $KN$ channel is always substantially below the $\pi\Sigma$ one. 
This prediction of the lowest derivatively coupled Lagrangian for the decay of a spin $\frac{3}{2}$ particle 
basically discards all experimental candidates to be $\Lambda_\text{S} \frac{3}{2}^-$. For example, for the $\Lambda(1520)$, the measurement for the ratio $\Gamma_{\pi\Sigma}/\Gamma_{KN}\sim 1$ whereas the prediction is a factor 9 (this is driven by the small phase space available for $KN$, but also on dynamical grounds).
While less extreme, the problem remains for the $\Lambda_\text{S}(1690)$ and higher reported candidates (less solid) with these quantum numbers. 
    
\subsection{Decay into three particles}\label{subsec:decay3}

Though data on three-body decays of excited hyperons are scant, it may be of interest
to think about them, at least for the one singlet candidate with the largest phase space,
the heavier $ \frac{1}{2}^+$. 
For these $\Lambda_\text{S} \rightarrow B + \phi +\phi$ decay processes, 
it is of note that two different $SU(3)$-singlet combinations can be formed in the final state with an octet baryon and two octet mesons, as
\begin{equation}
\textbf{8}\otimes \textbf{8}\otimes \textbf{8}= \textbf{1}\oplus \textbf{1} \oplus \hdots
\end{equation}
so that full specification of the final state requires a mixing angle
\begin{equation}
\arrowvert \psi_2 \phi_3 \phi_4\rangle = \cos \theta_M \arrowvert S_1\rangle +\sin \theta_M \arrowvert S_2\rangle
\end{equation}
which we have adopted, for this example, as $\theta_M=\frac{\pi}{4}$ (maximal mixing of the two singlets).

The construction of an appropriate chiral Lagrangian demands one of the mesons to be derivatively coupled, so that an appropriate effective vertex for a $\frac{1}{2}^+$ hyperon would be
    \cite{Thomas:2001} 
    \begin{equation}\label{Lag:3}
        L = \dfrac{\text{i}}{8 f_\pi^2} \left[\text{tr}(\overline{\Psi}_B \Phi \partial^\mu \Phi) +\text{tr}(\Phi \overline{\Psi}_B \partial^\mu \Phi )\right] \gamma_\mu \Psi_{\Lambda_\text{S}}
    \end{equation} where $f_\pi$ is the weak pion decay constant, $\Psi_{\Lambda_\text{S}}$ is the singlet baryon field, $\Psi_B$
the ground-state octet baryon and $\Phi$ the octet meson fields, respectively. The trace over the flavor index is sensitive to the ordering of the fields.

The tree-level matrix element, up to species-independent constants, follows 
from Eq.~(\ref{Lag:3})
to be \begin{eqnarray}
-\text{i}{\mathcal M} &\equiv&
\langle \psi_2 \phi_3 \phi_4 | {\mathcal L}_{\text{eff}} | \Lambda_\text{S} \rangle \nonumber \\ \nonumber &=& \left( \langle S_1\arrowvert \cos \theta_M + \langle S_2\arrowvert \sin \theta_M \right) {\mathcal L}_{\text{eff}} \arrowvert \Lambda_\text{S}\rangle \\
 &=& \frac{1}{\sqrt{2}}\overline{u}(2) \gamma_\mu (p_3^\mu + p_4^\mu) u(1)  /\sqrt{2} 
\end{eqnarray} 
from which
\begin{eqnarray} \label{Sq3body}
    \overline{|\mathcal{M}|^2} &= \nonumber &(2m_1^2 +m_3^2 +m_4^2 -m_{23}^2 -m_{24}^2 )(m_1^2 -m_4^2-m_{23}^2) \\
   & &-m_3 (m_{23}^2+m_{24}^2 -m_3^2 -m_4^2) +2m_1m_2 m_3^2
\end{eqnarray}
where $m_1$ is the mass of $\Lambda_\text{S}$ in center of mass frame, $m_2$ is the mass of the baryon, $m_3$ and $m_4$ are the mass of the mesons and, $m_{23}$ and $m_{24}$ are the Dalitz variables
\begin{equation}
    m_{23}^2 \equiv p_{23}^2 = (p_2+p_3)^2, \quad m_{24}^2 \equiv p_{24}^2 = (p_2+p_4)^2.
\end{equation}
A few more details on these variables, particularly to define the physical region over which the squared matrix element of Eq.~(\ref{Sq3body}) is integrated, are left for appendix~\ref{sec:Dalitz}.

A flavor-symmetry preserving decay of a flavor singlet hyperon can yield the particle combinations listed in table~\ref{tab:canales3}.
\begin{table}[h]
\caption{\label{tab:canales3}Three-body channels available for $SU(3)$-symmetry preserving decay of $\Lambda_\text{S}$. Their relative branching fractions with $J^\pi=\frac{1}{2}^+$ are given in figure~\ref{fig:3decay}}
\begin{tabular}{|c|c|c|c|c|c|c|c|c|c|c|}
     \hline
     Baryon & $\Sigma$ & $N$ & $\Xi$ & $\Lambda$ & $N$ & $\Sigma$ & $\Xi$ & $\Sigma$ & $\Lambda$ & $\Lambda$ \\ \hline
     $1^{\text{st}}$ Meson & $K$ & $\pi$ & $K$ & $K$ & $K$ & $\pi$ & $\eta$ &$\eta$ & $\eta$ & $\pi$ \\ \hline
     $2^{\text{nd}}$ Meson & $K$ & $K$ & $\pi$ & $K$ & $\eta$ & $\pi$ & $K$ & $\pi$ & $\eta$ & $\pi$ \\ \hline
\end{tabular}
\end{table}

\begin{figure}[h!]
\includegraphics[width=0.5\textwidth]{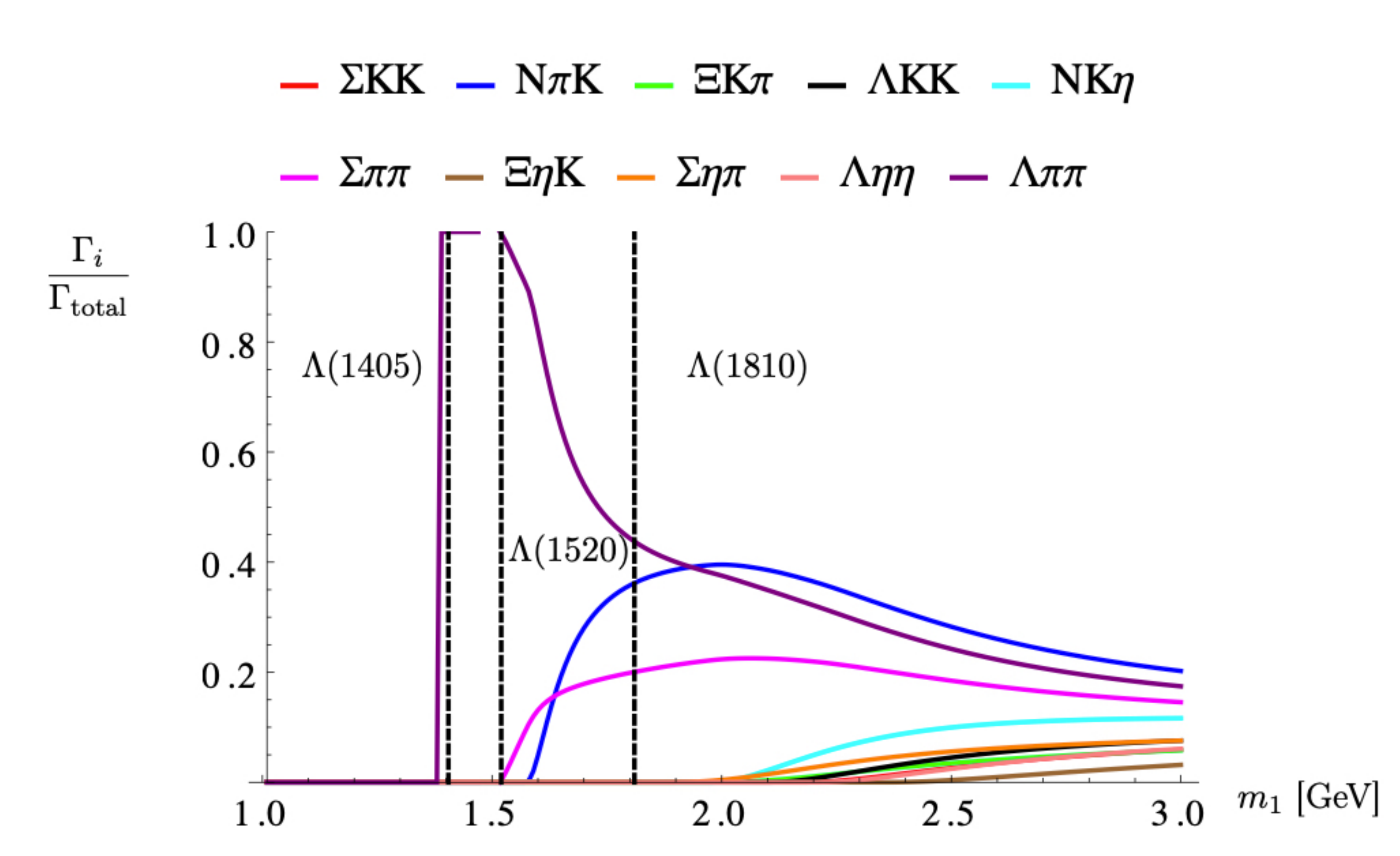}
\caption{
\label{fig:3decay}
Relative strength of the three-body decays in the limit of exact $SU(3)$ symmetry.}
\end{figure}

 The total width, and now also the branching ratios of these channels, are not rigorously accessible because the coupling constants in Eq.~(\ref{Lag:3}) and (\ref{decay2}) are not related in a model-independent way known to us, so that we cannot predict the ratio of three- to two- body decay fractions. What can be done is to once more exploit the symmetry structure built into Eq.~(\ref{Lag:3}) to predict the relative strength of three-body channels respect to each other. 
This relative strength $\Gamma_i/\Gamma(3\ {\rm body})$ is plot in figure~\ref{fig:3decay}.

To exemplify, let us for a moment take the $\Lambda(1810)$ as the lightest $\Lambda_\text{S} \frac{1}{2}^+$ candidate at face value, though a new data analysis suggests that it might be a surplus resonance not really necessary to obtain an optimal global data fit~\cite{Sarantsev:2019xxm}. No three-body decays seem to have been experimentally reported.

At the position of this $\Lambda(1810)$ (vertical line in figure~\ref{fig:3decay})
we see that $\Lambda\pi\pi$, the channel with the lowest threshold, starts losing its phase-space advantage, so that it is still dominant but comparable to $NK\pi$ (that becomes dominant for even higher masses) due to the derivative coupling of Eq.~(\ref{Lag:3}), and about a factor of 2 larger than $\Sigma\pi\pi$, with other decay channels being kinematically closed at that energy.

\section{Estimate of singlet mass in the NCState Coulomb--gauge QCD model: theoretical framework.}\label{sec:singletmass}

To attempt a numerical estimation of the actual $qqq$ singlet masses (as past work mostly took $uds$ configurations without a flavor separation), we employ a well-known  Hamiltonian model obtained from Coulomb gauge QCD. Its philosophy, dating back to~\cite{Robertson:1998va,LlanesEstrada:1999uh,Szczepaniak:2001rg}
is to use a field-theory formalism maintaining the global symmetries of QCD so that chiral symmetry breaking is spontaneous and not explicit as in the nonrelativistic quark model. The interaction corresponds to the established Cornell potential for heavy quarkonium with the appropriate color factors, and the flavor structure of the spectrum is reasonable; it can be seen as an extension of the Cornell model~\cite{Eichten:1979ms} to include gluodynamics. 

This approach gave a reasonable explanation of the lattice glueball spectrum~\cite{LlanesEstrada:2000jw,LlanesEstrada:2005jf}, basic features of quark-antiquark mesons and of three-quark baryons, and was deployed early on to show that $1^{-+}$ exotic mesons cannot be very light~\cite{LlanesEstrada:2000hj,General:2006ed}
 (unlike mainstream thought at the time); to study internal $b\bar{b}$ structure~\cite{TorresRincon:2010fu} abstracting model-independent features;
 and to study the coupling~\cite{Wang:2008mw} of $q\bar{q}$ and $qq\bar{q}\bar{q}$ configurations~\cite{General:2007bk}, with hints that this mixing would provide an explanation for ideal $\omega-\phi$ vector meson mixing.

The most recent works within the model's approach have been carried out by the Salvador de Bahia group~\cite{Abreu:2019adi,Abreu:2020ttf,Abreu:2020wio} in studying conventional $q\bar{q}$ spectroscopy in less trodden channels.

Thus, the model is a one-stop Hamiltonian for many issues in spectroscopy. On the down side, because it is an equal-time quantization approach, it is not useful to compute form factors or other functions pertaining to hadron structure, for which the Dyson-Schwinger+Bethe-Salpeter/Faddeev~\cite{Alkofer:2018yjm}, or the light-front~\cite{Choi:2017uos} or point form~\cite{Gomez-Rocha:2013zma} approaches are more apt.

The quark-part of the Hamiltonian is described in~\cite{LlanesEstrada:2004wr}
and contains a kinetic term, $H_{kin}$; the longitudinal Coulomb-potential interaction $V_C$ that accommodates asymptotic freedom at small distance and confinement at large distances; and an effective transverse interaction $V_T$ that represent the hyperfine quark-gluon interaction. It can be written in second quantization as
\begin{gather}
H = H_{kin} + V_C + V_T \label{h}\\ \noalign{\medskip}
H_{kin} = \int d^3\vec{x}\,\Psi^\dagger(\vec{x})(-i\vec{\alpha} \cdot \vec{\nabla} + m_f\cdot\beta)\Psi(\vec{x}) \label{hkin}\\ \noalign{\medskip}
V_C = -\frac{1}{2}\int d^3\vec{x}d^3\vec{y}\,\rho^a(\vec{x})V(|\vec{x} - \vec{y}|)\rho^a(\vec{y}) \label{vc}\\ \noalign{\medskip}
V_T = \frac{1}{2}\int d^3\vec{x}d^3\vec{y}\,J^a_i(\vec{x})\left(\delta_{ij} - \frac{\nabla_i\nabla_j}{\nabla^2}\right)U(|\vec{x} - \vec{y}|)J^a_j(\vec{y}) \ . \label{vt}
\end{gather}

Therein the quark fields $\Psi$ are used to construct a local color density $\rho^a$ and current $\vec{J}^a(\vec{x})$ with the color Gell-Mann matrices  $T^a$,
\begin{equation}\label{rhoAndJ}
\rho^a(\vec{x}) = \Psi^\dagger(\vec{x})T^a\Psi(\vec{x}), \quad \vec{J}^a(\vec{x}) = \Psi^\dagger(\vec{x})\vec{\alpha}T^a\Psi(\vec{x}) \ .
\end{equation}

The kernel $V$ has been presented in \cite{Szczepaniak:2001rg} as
\be\label{V}
V(q)=\begin{cases}
C(q) = -\frac{8.07}{q^2}\frac{\log^{-0.62}\left(\frac{q^2}{m^2_g} + 0.82\right)}{\log^{0.8}\left(\frac{q^2}{m^2_g} + 1.41\right)},
& \mbox{for $q > m_g$}\\ 
L(q) = -\frac{12.25m^{1.93}_g}{q^{3.93}}, & \mbox{for $q < m_g$}
\end{cases}
\ee
\\
$q$ is the modulus of the exchanged momentum. The model depends on a free parameter, the dynamical mass of the exchanged gluon, that   takes a value\cite{LlanesEstrada:2004wr} $m_g = 0.6$ GeV, yielding an asymptotic string tension-like scale of $\sqrt{8\pi\sigma}~12.25^{1/1.93}m_g \simeq 0.44 \rm GeV$, that is, $\sigma \simeq 0.2 {\rm GeV}^2$ sufficient for a reasonable description of the quarkonium spectrum. \\

The kernel $U$ is introduced as {\it model 4} in \cite{LlanesEstrada:2004wr}. It is a Yukawa-type potential of the form:
\begin{equation}\label{U}
U(q)=\begin{cases}
C(q), & \mbox{for $q > m_g$}\\
-\frac{C_h}{q^2 + m^2_g}, & \mbox{for $q < m_g$} \ .
\end{cases}
\end{equation}
\\
(The constant      $C_h\simeq 
(2.907{\rm GeV})^3$
for $m_g=0.6$ GeV simply guarantees continuity of $U(q)$ at the matching point).
 
It is a natural implementation of the Coulomb gauge philosophy that separates an infrared strong scalar potential and  an infrared suppressed transverse one due to physical gluon exchange being affected by the dynamical mass $m_g$. \\

The quark field can be expanded in particle/antiparticle normal modes in momentum space,
\begin{equation}\label{field}
\Psi(\vec{x}) = \int \frac{d^3\vec{k}}{(2\pi)^3}e^{i\vec{k}\vec{x}}\sum_{\substack{\lambda c f}}\left({\cal U}_{\vec{k}\lambda f}B_{\vec{k}\lambda f c} + {\cal V}_{-\vec{k}\lambda f}D^\dagger_{-\vec{k}\lambda f c}\right)\hat{\epsilon}_c\hat{\eta}_f 
\end{equation}\\

$\lambda, f$ and $c$ are indices for helicity, flavor and color respectively; and $\hat{\epsilon}_c$, $\hat{\eta}_f$ are the color and flavor unitary vectors. The spinors ${\cal U}, {\cal V}$ are, in terms of the Pauli spinors ($\chi_\lambda$), given by
\be
{\cal U}_{\vec{k}\lambda f} = \frac{1}{\sqrt{2}}\left[\begin{array}{c}\sqrt{1+s_{kf}}\ \chi_\lambda \\ \sqrt{1-s_{kf}}\ \vec{\sigma}\cdot\hat{k}\chi_\lambda
\end{array}\right] 
\ee 
\be
{\cal V}_{-\vec{k}\lambda f} = \frac{1}{\sqrt{2}}\left[\begin{array}{c}-\sqrt{1-s_{kf}}\ \vec{\sigma}\cdot\hat{k}i\sigma_2\chi_\lambda \\ \sqrt{1+s_{kf}}\ i\sigma_2\chi_\lambda
\end{array}\right]
\ee
\\
Where we have made use of the Bogoliubov angle, $\phi_{kf}$, related to a running quark mass $m(k, f)$ and energy $E(k, f) = \sqrt{m^2(k, f) + k^2}$ in the following way
\begin{equation}\label{sincos}
s_{kf} = {\rm sin}\ \phi_{kf} = \frac{m(k, f)}{E(k, f)}, \quad c_{kf} = {\rm cos}\ \phi_{kf} = \frac{k}{E(k, f)}
\end{equation}
\par

\begin{figure}[h]
\includegraphics[width=\columnwidth]{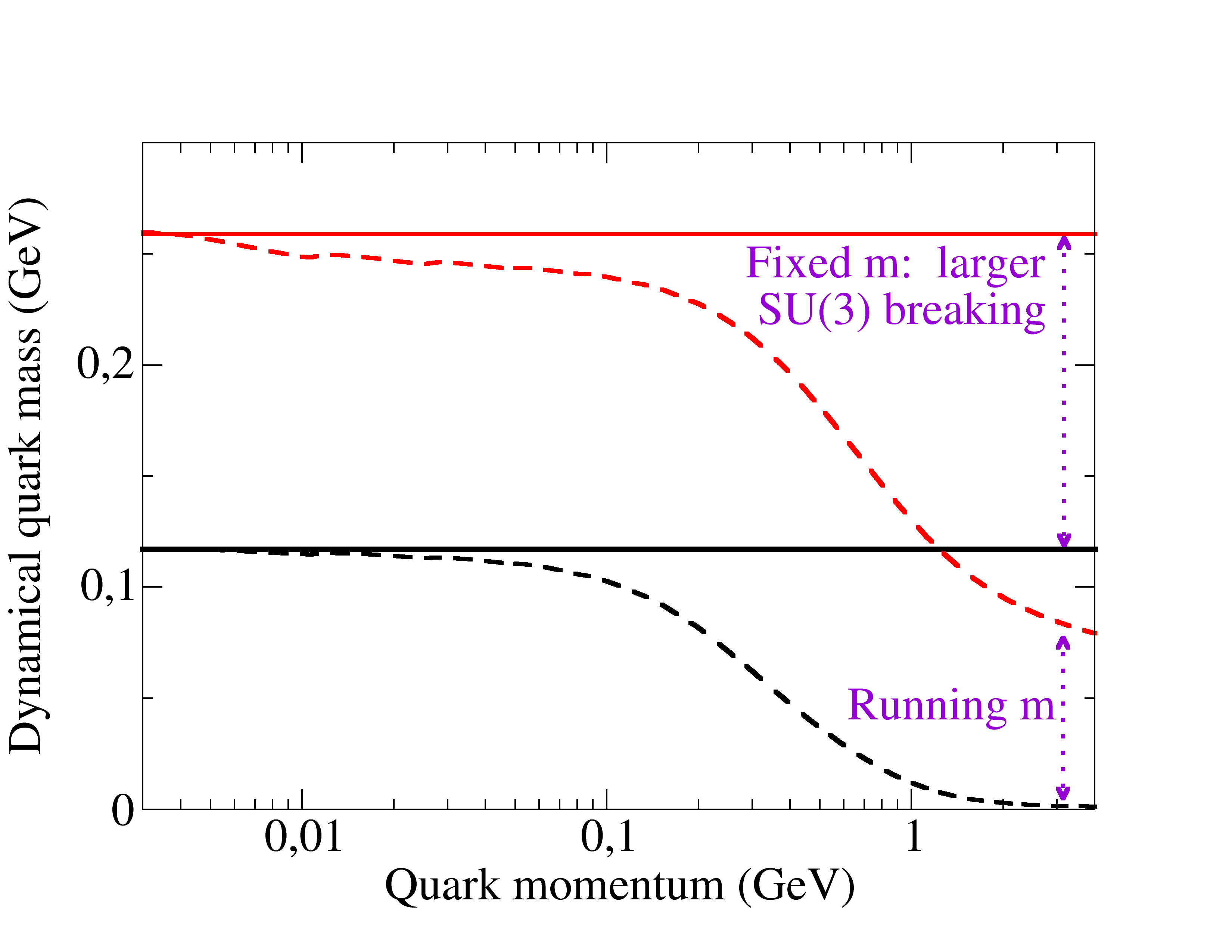}
\caption{\label{fig:gapeq} Typical running quark mass $m(k)$ in the one-particle spinors solving the Hamiltonian model gap equation for the light and strange sector. The $SU(3)$-breaking scale $m_s-m_u\simeq 70$ MeV at a high quark momentum is enhanced by dynamical chiral symmetry breaking and becomes a factor 2 larger at low momentum in this particular calculation. The constituent quark model, with a fixed quark mass instead, breaks $SU(3)$ with the same (larger) intensity at all scales.}
\end{figure}

The gap equation that provides the model vacuum and one-particle dispersion relation was reported in an earlier work~\cite{LlanesEstrada:2004wr}. In figure~\ref{fig:gapeq} we plot a couple of typical $m(k)$ mass functions for quark momentum up to a few GeV. The generated quark mass seems somewhat smaller than usual in the constituent picture, but this is not remarkable in an approach where there is a significant self-energy in the potential part of the Hamiltonian (see equation~(\ref{final_mass}) below). 

It is clear that $SU(3)$ symmetry breaking by the effective quark mass is largest for zero momentum quarks and drops with the scale (just as it should in exact QCD). This leads us to expect less $SU(3)$ symmetry breaking (and therefore, less $\Lambda_\text{S}$-$\Lambda_O$ singlet-octet mixing) than in constituent quark models:
those feature a constant quark mass which is scale-independent, and therefore the high-momentum wavefunction components support larger flavor-symmetry breaking.

Now with all these pieces and shortening $B_i = B_{\vec{k}_i\lambda_i f_i c_i}$ for the $i$th quark, we can write down the state of our singlet baryon $|\Lambda_\text{S}\rangle$ with well defined $J^\pi$ in terms of a suitable  combination of products of the spatial ans\"atze of each quark $F^{\lambda_1\lambda_2\lambda_3}_{\Lambda_\text{S}}(\{\vec{k}_i\})$ as
\be
\begin{split}\label{lambdas}
|\Lambda_\text{S}\rangle = \int \frac{d^3\vec{k}_1d^3\vec{k}_2d^3\vec{k}_3}{(2\pi)^9}\frac{\epsilon^{c_1c_2c_3}}{\sqrt{6}}\frac{\epsilon^{f_1f_2f_3}}{\sqrt{6}}\delta(\vec{k}_1 + \vec{k}_2 + \vec{k}_3)\times \\F^{\lambda_1\lambda_2\lambda_3}_{\Lambda_\text{S}}(\{\vec{k}_i\})
B^\dagger_1B^\dagger_2B^\dagger_3|\Omega\rangle,
\end{split}
\ee
(In Eq. (\ref{lambdas}), summations over helicity, flavor and color are implicit.)\\ 

Hence, we can express the variational approximation to the $\Lambda_\text{S}$ mass as

\begin{widetext}
\begin{multline}
\label{final_mass}
M_{\Lambda_\text{S}} = \langle \Lambda_\text{S}|H|\Lambda_\text{S}\rangle = 3\int \frac{d^3\vec{k}_Ad^3\vec{k}_B}{(2\pi)^6}\left[{\cal F}^{\lambda_A\lambda_B\lambda_C}_{\Lambda_\text{S}}(\vec{k}_A, \vec{k}_B)\right]^\dagger \Bigg\{\sum_{f_A}\frac{c_{k_Af_A}|\vec{k}_A| +  m_{f_A}s_{k_Af_A}}{3}{\cal F}^{\lambda_A\lambda_B\lambda_C}_{\Lambda_\text{S}}(\vec{k}_A, \vec{k}_B)   
-\frac{2}{3} \frac{d^3\vec{q}}{(2\pi)^3}\times \\ \noalign{\medskip} \Bigg(V(|\vec{q}|)\bigg[{\cal F}^{\lambda_A\lambda_B\lambda_C}_{\Lambda_\text{S}}\left(\vec{k}_A, \vec{k}_B\right)\sum_{f_A}\frac{1}{3}(s_{k_Af_A}s_{k_A + q, f_A} + c_{k_Af_A}c_{k_A + q, f_A}\cdot x)
-\frac{1}{6}\sum_{\substack{f_Af_B\\f_A \neq f_B}}{\cal U}^\dagger_{\vec{k}_A\lambda_Af_A}{\cal U}_{\vec{k}_A + \vec{q}, \lambda_af_A}{\cal U}^\dagger_{\vec{k}_B\lambda_Bf_B}{\cal U}_{\vec{k}_B - \vec{q}, \lambda_b f_B}\times \\ \noalign{\medskip}{\cal F}^{\lambda_a\lambda_b\lambda_C}_{\Lambda_\text{S}}\left(\vec{k}_A + \vec{q}, \vec{k}_B - \vec{q}\right)\bigg]
+ U(|\vec{q}|)\bigg[\sum_{f_A}\frac{1}{3}\left(2s_{k_Af_A}s_{k_A + q, f_A} + 2c_{k_Af_A}c_{k_A + q, f_A}\frac{x(k_A^2 + (\vec{k}_A + \vec{q})^2) - |\vec{k}_A + \vec{q}|k_A(1 + x^2)}{q^2}\right)\times\\ \noalign{\medskip}{\cal F}^{\lambda_A\lambda_B\lambda_C}_{\Lambda_\text{S}}\left(\vec{k}_A, \vec{k}_B\right) 
- \sum_{\substack{f_Af_B\\f_A \neq f_B}}{\cal U}^\dagger_{\vec{k}_A\lambda_Af_A}\alpha_i\ {\cal U}_{\vec{k}_A + \vec{q}, \lambda_af_A}\frac{\left(\delta_{ij} - \hat{q}_i\hat{q}_j\right)}{6}{\cal U}^\dagger_{\vec{k}_B\lambda_Bf_B}\alpha_j\ {\cal U}_{\vec{k}_B - \vec{q}, \lambda_b f_B}{\cal F}^{\lambda_a\lambda_b\lambda_C}_{\Lambda_\text{S}}\left(\vec{k}_A + \vec{q}, \vec{k}_B - \vec{q}\right)\bigg]\Bigg)\Bigg\}    
\end{multline}
\end{widetext}

Where we have employed the usual shorthand $x = \frac{\vec{k}_A}{|\vec{k}_A|}\cdot \frac{\vec{k}_A + \vec{q}}{|\vec{k}_A + \vec{q}|}$ . Also, notice the difference among $F_{\Lambda_\text{S}}$ in Eq. (\ref{lambdas}) and ${\cal F}_{\Lambda_\text{S}}$ in Eq. (\ref{final_mass}): the first is the product of the spatial Ans\"atze of the 3 quarks, while the last is its antisymmetrized form
\be
{\cal F}_{\Lambda_\text{S}}^{\lambda_1 \lambda_2 \lambda_3} = \sum_{a, b, c}\epsilon^{abc} F_{\Lambda_\text{S}}^{\lambda_a \lambda_b \lambda_c}(\vec{k}_a, \vec{k}_b, \vec{k}_c),\ \{a, b, c\} \in \{1, 2, 3\}
\ee
\\
This fermion antisymmetry naturally appears due to the anticommutation rules of the creation and annhilation operators $B^\dagger_i, B_i$.

It remains to specify the parameters of the Hamiltonian. They are consistent with extensive meson work in the Coulomb gauge model, but also with earlier baryon computations that addressed multiple spin nucleon resonances in search for parity doublets, and are discussed in table~\ref{tab:params}.
\begin{table}[h!]
    \centering
    \begin{tabular}{|c|c|}
         \hline
         Hamiltonian parameters &\\
         \hline\hline
         $m_u=m_d$ & 0.001 GeV \\
         \hline
         $m_s$ & 0.070 GeV\\
         \hline
         $m_g$ & 0.6 GeV\\
         \hline\hline
         Integration controls &\\
         \hline\hline
         $\lambda_{IR}$ & $3\cdot10^{-3}$\,GeV \\ \hline
         $\Lambda_{UV}$ & $\propto$ variational parameters\\  \hline
    \end{tabular}
    \caption{Parameters used in the model Hamiltonian and in the integration. The strange $m_s$ is set at 70 MeV but we also perform runs at 25 MeV to check dependence thereof, and both are fixed in the gap function (momentum scheme) around 2 GeV; and $m_g$ controls the kernels $V,\ U$ of Eq.~(\ref{V}) and~(\ref{U}), with the interpretation of a longitudinal gluon mass-like parameter. The integral extends between an IR cutoff (to avoid an accidental divergence in the Monte Carlo with the IR-strong potential, but there is no dependence in it) and an UV cutoff to cover (most of) the corresponding variational wavefunction. 
    \label{tab:params}}
\end{table}

The two (current) quark masses are near actual parameters in the QCD Lagrangian at 2 GeV, the reason being the implementation of spontaneous chiral symmetry breaking by a gap equation, unlike constituent quark models. The effective gluon mass present in the kernel on Eq. (\ref{V}) was set to $m_g=0.6$\,GeV. 
Those parameters are fixed from the meson sector of the theory, and yield around $m_\pi = 150$ MeV. 
Because spontaneous symmetry breaking is implemented, Goldstone's theorem and the Gell-Mann-Oakes-Renner relation are satisfied, so in the chiral limit $m_\pi=0$; fine tuning $m_u$ easily yields the physical pion mass, but we see no point in reaching such precision.
As for the basic vector meson, with this set $m_{\rho/\omega} = 730$ MeV  (about 40 MeV too low, but this resonance is 150 MeV broad, so this is not a big deal numerically). Finally, $m_\phi$ (a pure $s\bar{s}$ meson) has a mass of 1030  MeV (again quite acceptable as its physical mass is 1020 MeV) for that value $m_s=70$ MeV.

\section{Estimate of singlet mass in the NCState Coulomb--gauge QCD model: extensive numerical computations.}\label{sec:numeric}

In order to compute the mass of the SU(3) flavor singlet, we have to evaluate the matrix elements of the Hamiltonian (presented in section~\ref{sec:singletmass}), with each of the selected families of variational wave functions. For that, the complete theoretical framework was implemented in a C++ program where the $\vec{k}_{A}$, $\vec{k}_{B}$ and $\vec{q}$ momentum integrals of Eq.~(\ref{final_mass}) were estimated using the Monte Carlo-based multi-dimensional Cuba library~\cite{Hahn:2004fe.Cuba}. Most frequently, we employed the well-known Vegas algorithm therein~\cite{Lepage:1980dq},   though we have also cross checked with some of the other integration algorithms in the package.

The color Ward identities between the gap equation and the $qqq$ kernel guarantee infrared finiteness~\cite{Bicudo:1989si,LeYaouanc:1984ntu} of the matrix element in Eq.~(\ref{final_mass}) with the employed potential, that in the infrared  $V \propto q^{-4+\epsilon}$ practically is the Fourier transform of a linear confining kernel.

Still, the nine-dimensional momentum integral was regulated with an IR-cutoff of $3\cdot10^{-3}$\,GeV to avoid any accidental apparent divergence, particularly in the exchanged momentum, $q$ in Eq.~(\ref{final_mass}), due to the random distribution of points in the Monte Carlo algorithm.

Additionally, for the Monte Carlo algorithm to correctly cover most of each variational wavefunction, an upper integration limit ($\Lambda_{UV}$) was introduced. As each wave function extends to different maximum momentum, this integration cutoff is scaled as a multiple of the (inverse) variational parameter. Therefore, it takes a different value in each of the  computations, typically of order 3-10 times the relevant scale. For example, in the next subsection~\ref{testcode}, the quoted values were obtained with $\Lambda_{UV}=5\times {\rm MAX}(v_\rho,v_\lambda)$ as described therein.

That a small tail of the nominal wavefunction may extend outside the integration domain (and failed to be integrated over) does not cause a problem of principle: it amounts to a redefinition of the  variational wavefunction as including an additional truncation parameter, so that it is multiplied by a step function. (There are smaller orthogonalization effects that need not concern us at the level of precision that the Monte Carlo evaluation achieves.) This is legitimate within the variational principle, as long as the same truncation of the integration is applied to the normalization so that $\frac{\langle \Lambda_\text{S} \arrowvert \Lambda_\text{S} \rangle}{\langle 0 \arrowvert 0 \rangle} = 1$. Therefore, we compute the normalization of the wavefunction with the same computer code and cutoffs, then use the obtained number to set it to 1. The variational approach is then sensible in spite of cutting off the integrations.

\subsection{Computer code test: $\Delta$($\frac{3}{2}^+$) and $\Omega$($\frac{3}{2}^+$) baryons} \label{testcode}
Prior baryon computations in this scheme~ \cite{Bicudo:2009cr,Bicudo:2016eeu,LlanesEstrada:2011jd} 
focused on neutron-wavefunction anisotropic deformation under the high compression of neutron stars and on parity doubling in the highly excited $N/\Delta$
spectrum.

As a renewed test of the computer code, modified for this singlet hyperon application, the masses of two well known baryons, Gell-Mann's decuplet $\Delta(1232)$ and $\Omega(1672)$, were computed first. These two baryons are archetypical $qqq$ states with three light quarks (the $\Delta$) and three strange quarks (the $\Omega$), having particularly simple, completely symmetric $qqq$ wavefunctions. We assume here perfect isospin symmetry so $m_u=m_d$. They are thus ideal cases to test the entire computer program (except, of course, the singlet wavefunction construction).

The parameters of these two Hamiltonian computations have been shown on Table \ref{tab:params}, and they are consistent with meson and earlier baryon work in the same model.

For this calibration exercise, variational wave functions were adapted from \cite{LlanesEstrada:2011jd,Bicudo:2009cr,Bicudo:2016eeu}. The radial wave function takes a rational form,
\begin{equation}
    R(k) = \frac{1}{\left[\left(\frac{|\bf k_\rho|}{v_\rho}\right)^4+1\right]\left[\left(\frac{|\bf k_\lambda|}{v_\lambda}\right)^4+1\right]} 
    \label{rationalWF}
\end{equation}
with 
\begin{eqnarray}
{\bf k}_\rho &:=& \frac{1}{\sqrt{2}}({\bf k}_1-{\bf k}_2) \nonumber \\
{\bf k}_\lambda &:=& \sqrt{\frac{3}{2}}({\bf k}_1+{\bf k}_2)\propto ({\bf k}_1+{\bf k}_2-2{\bf k}_3)
\end{eqnarray}
appropriate Jacobi coordinates for the three body problem in the center of momentum frame in which 
\begin{equation}
{\bf k}_1+{\bf k}_2+{\bf k}_3={\bf 0}\ .
\end{equation}
Each of their moduli is scaled by a corresponding $v_\rho$ and $v_\lambda$ variational parameter. This two-dimensional parameter space will later be scanned for a minimum of the variational mass, according to the Rayleigh-Ritz variational principle.

\begin{figure}[h!]
    \centering
    \includegraphics[width=\columnwidth]{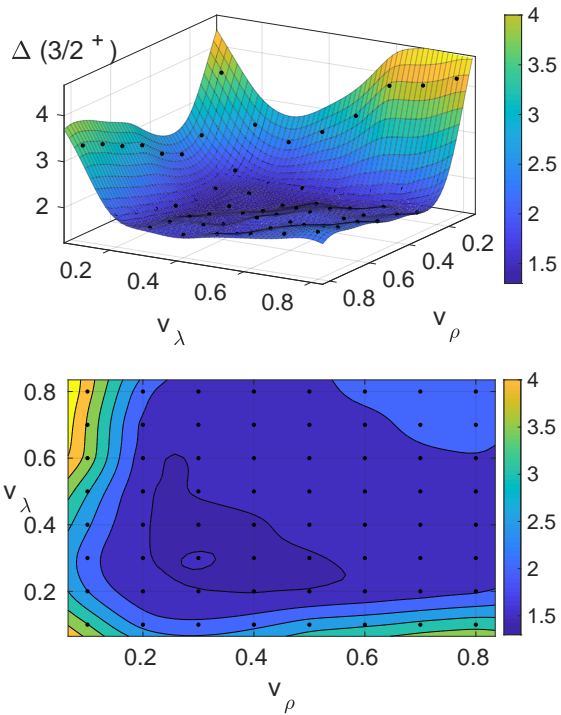}
    \caption{$\langle H \rangle$ over the  variational parameter space of the $\Delta(1232)$ calibration test. The wave function (of rational form) is that of Eq.~(\ref{rationalWF}). The minimum  energy obtained on the discrete grid is given on Table \ref{DeltaOmegaMinimum}.
    For visualization, a continuous surface is obtained from a bi-harmonic spline interpolation of the discrete values result of the MC simulations.  }
    \label{delta}
\end{figure}

\begin{figure}[h!]
    \centering
    \includegraphics[width=\columnwidth]{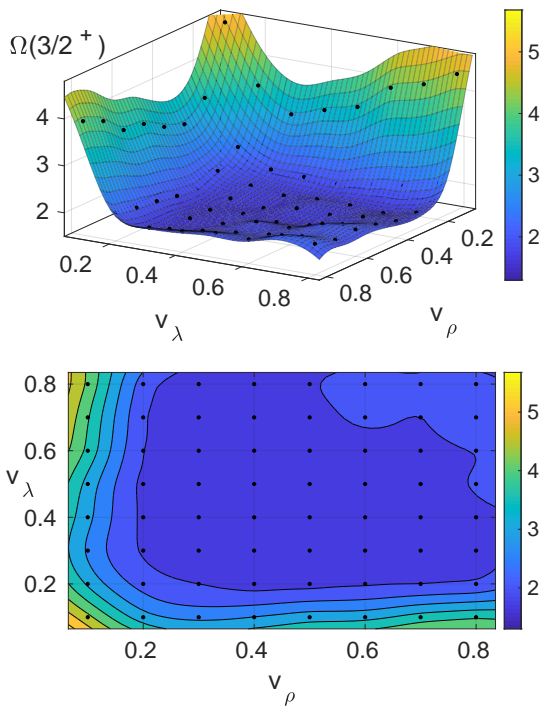}
    \caption{$\langle H \rangle$ over the  variational parameter space of the $\Omega(1672)$ calibration test. The same wave function and methodology as in the $\Delta(1232)$'s exercise of figure~\ref{delta} was used. The minimum of $E(v_\rho,v_\lambda)$ over the discrete grid is given in Table \ref{DeltaOmegaMinimum}.}
    \label{omega}
\end{figure}

The variational parameter spaces for the $\Delta(1232)$ and $\Omega(1672)$ are explored in Figures \ref{delta} and \ref{omega} respectively. For each parameter pair in the two--dimensional grid, we obtained $\langle H\rangle(v_\rho,v_\lambda)$. After this calculation of $E$ over the variational-parameter space grid, we extract its minimum that, by the variational principle, is an upper bound to the respective ground level energy. Those minima are carried over to Table \ref{DeltaOmegaMinimum}. 

The discrete values of $E$ were continuously interpolated by a bi-harmonic spline for better visibility in the two figures. As is usual in these calculations, when the energy is known to precision $\epsilon$, the wavefunction parameter is only obtainable to precision $\sqrt{\epsilon}$ (since the matrix element is quadratic in the wavefunction). Therefore, the minima present themselves as broad valleys, depicted with  the darkest shades in figures~\ref{delta}, \ref{omega} (and following). In those dark areas, values of $\langle H \rangle $ under 2 GeV are found.

\begin{table}[h!]
\caption{Calibration computation of the $\Delta(1232)$ (all light quarks) and $\Omega^-(1672)$ (three strange quarks), with maximal spin $ J=S= \frac{3}{2}$, $L=0$. The $\Delta$ comes out  $150$ MeV heavier than the datum, in line with expectations for a pure $qqq$ computation that does not incorporate its $\pi N$ channel. Since it has a 130 MeV width, a positive 150 MeV deviation is very reasonable for this variational computation. Gell-Mann's  $\Omega^-$ state, stable by the strong force, is calculated compatibly with its experimental mass. A few values around the minimum,  rounded off to 5 MeV precision, are quoted.\label{DeltaOmegaMinimum}}
\begin{tabular}{|c|c|c|c|}
\hline
  & \quad$v_\rho$\quad &  \quad $v_\lambda$ \quad  &\quad $\langle H \rangle $ [GeV] \quad\\ \hline \hline
 & 0.3 & 0.4 & $\bf  1.38 \pm 0.04 $\\ \cline{2-4}
\quad $\Delta(1232)$ \quad & 0.4 & 0.4 & $1.46 \pm 0.05 $\\ \cline{2-4}
 & 0.4 & 0.3 & $1.45 \pm 0.04 $\\ \hline \hline
 & 0.4 & 0.4 & $\bf 1.65 \pm 0.04 $\\ \cline{2-4}
\quad $\Omega(1672)$ \quad & 0.4 & 0.3 & $1.73 \pm 0.03 $\\ \cline{2-4}
 & 0.5 & 0.3 & $1.69 \pm 0.04 $\\ 
 \cline{2-4} & 0.5 & 0.4 & $1.72 \pm 0.05 $\\ 
 \hline
\end{tabular}
\end{table}

According to the Rayleigh-Ritz variational principle, all energies calculated are  upper bounds to the physical particle mass within the given Hamiltonian, with the optimal one corresponding to the minimum of the variational surface. Nevertheless, due to the Monte Carlo computational method, down-fluctuations can occur.
Thus, we strove to increase the number of integration points until the number of fluctuations was small enough to keep the standard deviation at or below the 50 MeV level. The Monte Carlo uncertainty was reduced to this level as shon in table~\label{DeltaOmegaMinimum} (the uncertainty quoted there corresponds only to this Monte Carlo computation of the matrix element, and not to the error induced by the variational principle, whose sign is known, but not its size).

The computer code was run at the modest group cluster of the theoretical physics department in Madrid  and similar facilities.

The $\Omega(1672)$ computed energy is in fair agreement with the experimental value, indicating a correct implementation of the strange quark framework.  
The $\Delta(1232)$ comes out $\sim 150$ MeV heavier than its physical mass, in line with expectations for a pure $qqq$ computation that does not incorporate its $\pi N$ channel. Since it has a 130 MeV width, a positive 150-200 MeV deviation is very reasonable for a variational computation.

From these calibration tests we take the accuracy of the computations within the Coulomb-gauge Hamiltonian, including  the Monte Carlo integration procedure, as validated,  and reassert the adequacy of the Hamiltonian  parameters used in past computations. 

\subsection{$\Lambda_\text{S}(qqq)$ flavor-singlet mass computation}

We then proceed to the goal of this section.
The only changes needed to determine the $\Lambda_\text{S}$ states masses with the same Hamiltonian tested in subsection~\ref{testcode} concern the variational wavefunctions. 
Given the reasonable performance with the two tested single-flavor baryons in the decuplet, only a few modifications concerning the multi-flavor structure had to be made. 

Since their symmetry is more complicated, employing two variational parameters for the Jacobi variables $k_\rho$ and $k_\lambda$ turned out not to be the most straight-forward procedure. Instead, the wavefunction was written down for each quark (guaranteeing the correct symmetry by applying appropriate symmetrization/antisymmetrization operators) so that three variational parameters had to be used. In comparing with the one-flavor cases, the mixed wave functions required a significantly large amount of computational time. To reduce it, we used the  anti-symmetry of the wave functions (explicitly tested in the code) to reduce the number of computed points in the variational space. This feature allowed us to speed the computation by a factor $\frac{N^N}{N!}$. 

\begin{figure}[h!]
    \centering
    \includegraphics[width=\columnwidth]{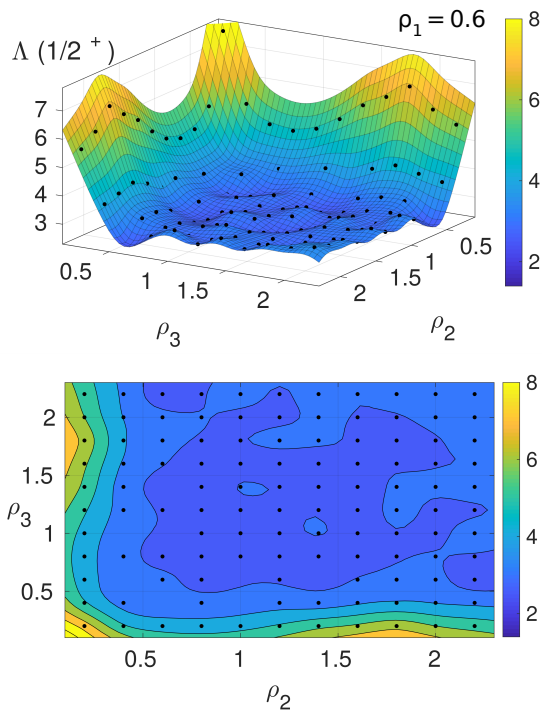}
    \caption{ 
    $\langle H \rangle$ over the  variational parameter space for the  $\Lambda(\frac{1}{2}^+)$ with hydrogen-like wavefunction $R^{(0)}$. The first variational parameter $\rho_1$ was fixed to 0.6 for which the lowest values where found. As in the calibration tests, the direct output of the Monte Carlo simulations was fitted to a continuous surface through a bi-harmonic spline interpolation. The minima providing the optimal upper bound for the energy of the baryon following the  Rayleigh-Ritz principle are given in Table \ref{HydrogenLambdaS}.}
    \label{JP0}
\end{figure}

\begin{figure}[h!]
    \centering
    \includegraphics[width=\columnwidth]{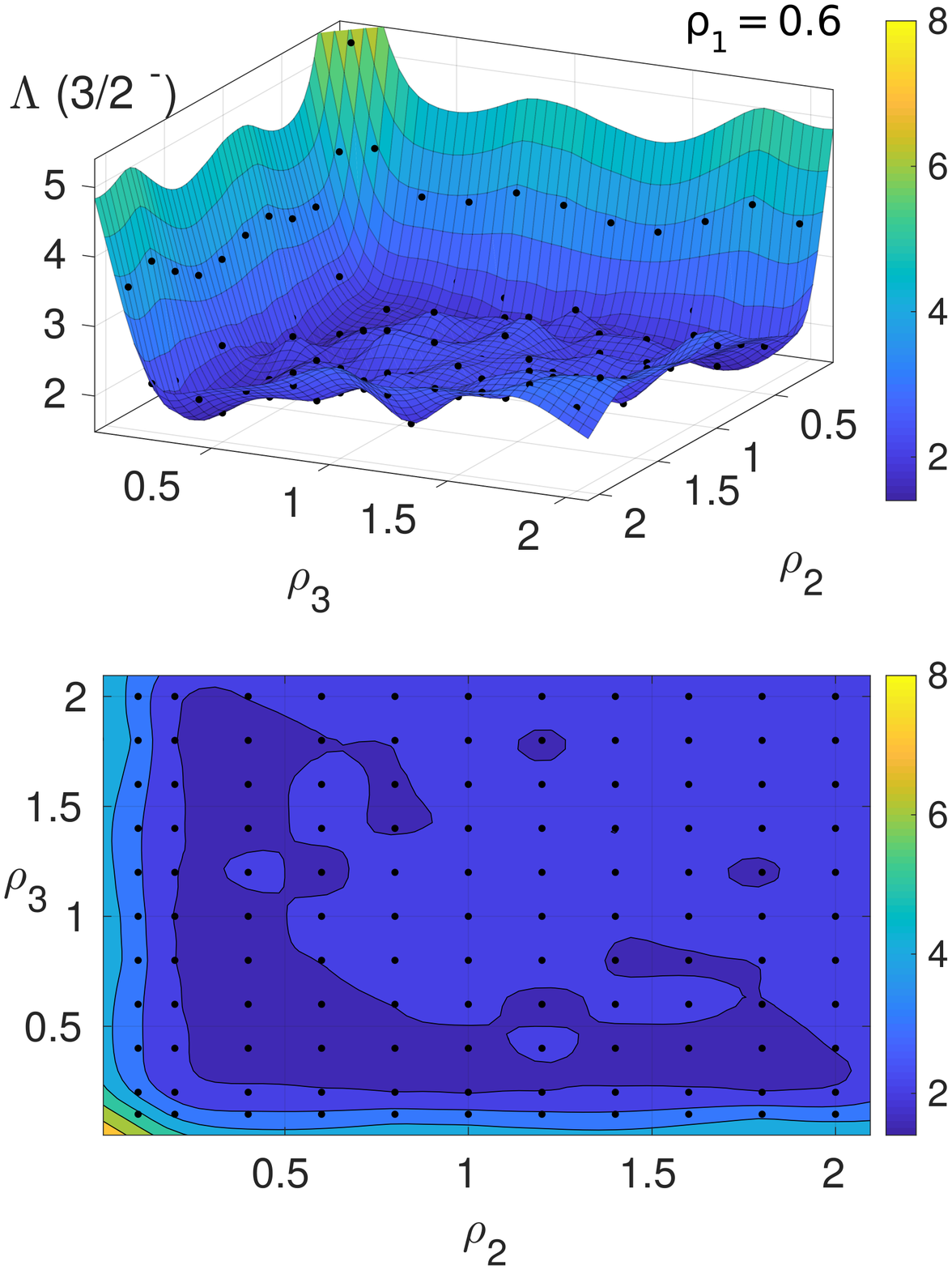}
    \caption{Same as Figure \ref{JP0} for the $\Lambda(\frac{3}{2}^-)$ state, with hydrogen-like wavefunction $R^{(0)}$. }
    \label{JP1}
\end{figure}

\begin{figure}[h!]
    \centering
    \includegraphics[width=\columnwidth]{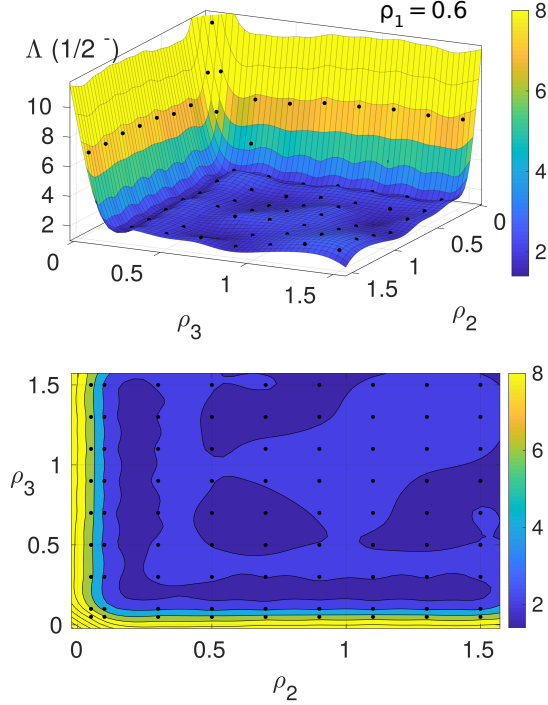}
    \caption{Same as Figure \ref{JP0} for the $\Lambda(\frac{1}{2}^-)$ state with
    hydrogen-like variational wavefunction $R^{(0)}$.}
    \label{JP2}
\end{figure}

We have run the computer codes with all four 
radial Ans\"atze in Eq.~(\ref{radial0}) and following. 
The variational principle indicates that in each channel we should keep the minimum energy over each Ansatz family, and then in turn select the minimum among the four families.
Figures~\ref{JP0}, \ref{JP1} and \ref{JP2} show $\langle H \rangle$ over the variational space for $R^{(0)}$, the variational wavefunction that is hydrogenlike, one figure for each $J^\pi$.

 That hydrogen-like $R^{(0)}$ and $R^{(3)}$, based on an interpolation to the solution of the previously computed two-body problem, were generally superior to both of the harmonic oscillator (one or three-dimensional) wavefunctions $R^{(1)}$ and $R^{(2)}$. For the $\frac{1}{2}^-$ and  $\frac{3}{2}^-$, we quote results from the hydrogen-like $R^{(0)}$, that was as good as any (see table~\ref{tab:parametrosvar}) and is of simple physical interpretation.

\begin{table}[h]
\centering
\begin{tabular}{|c|c|c|c|c|}
    \hline
     \multicolumn{5}{|c|}{Hydrogen-like Ansatz for $\Lambda_\text{S}$}\\
     \hline
      \hline
     $J^\pi$ & $\rho_1$ [GeV] & $\rho_2$  [GeV] & $\rho_3$ [GeV] & Mass (GeV) \\
     \hline \hline
     
    & 0.6 & 0.8 & 1.1 & $2.7 \pm 0.2$ \\
    \cline{2-5}
    & 0.6 &	0.8 & 1.2 &	$2.7 \pm 0.3$ \\
    \cline{2-5}$\frac{1}{2}^+$
    & 0.6 & 0.8 & 1.3 &	$2.8 \pm 0.2$ \\
    \cline{2-5}
    & 0.6 & 0.8 & 1.4 &	$2.6 \pm 0.2$ \\
    \cline{2-5}
    & 0.6 & 1.0 & 1.0 & $2.8 \pm 0.2$ \\
    \hline\hline
    
    & 0.4 & 0.6 & 1.4 & $1.8 \pm 0.2$ \\
    \cline{2-5}$\frac{3}{2}^-$
    & 0.4 & 0.6	& 1.6 &	$1.7 \pm 0.2$ \\
    \cline{2-5}
    & 0.4 & 0.8 & 0.8 &	$1.6 \pm 0.3$ \\

    \hline\hline
    
    
    & 0.4 &	0.4 & 1.4 &	$1.9 \pm 0.1$ \\
    \cline{2-5}$\frac{1}{2}^-$
    & 0.4 &	0.4 & 1.6 &	$1.8 \pm 0.2$ \\
    \cline{2-5}
    & 0.4 &	0.6	& 0.6 &	$2.0 \pm 0.1$ \\
    \hline
\end{tabular}
\caption{$\Lambda_\text{S}$ mass values around the minimum of each $J^\pi$ state and hydrogen-like radial wavefunction ($R_n^{(0)}(k)$ ansatz, Eq.~(\ref{radial0}))
\label{tab:parametrosvar}}
\end{table}

For $\frac{1}{2}^+$ instead,  the value of 2.7 GeV quoted in table~\ref{tab:parametrosvar}, being a GeV above the singlets in the other channels, looks unnatural to us. 
In this case we found the tabulated, interpolated and rescaled $R^{(3)}$ to be optimal: the minimum of $\langle H \rangle$ drops by 0.3 GeV respect to the hydrogen-like $R^{(0)}$ to yield the $2.4$ GeV quoted in table~\ref{HydrogenLambdaS}.
That table collects the optimal value that we have been able to locate for each $J^\pi$ combination and is the final result of this section.
\begin{table}[]
\centering
\begin{tabular}{|c|c|c|c|c|}

\hline
     \multicolumn{5}{|c|}{Minimum $\langle H\rangle_{\Lambda_\text{S}}$ with meson-derived Ansatz $R^{(3)}$}\\
     \hline
     \hline 
     $J^\pi$ & $\rho_1$ & $\rho_2$ & $\rho_3$ & $E_0$ [GeV] \\
     \hline \hline
     
     $\frac{1}{2}^+$ & $2.5 \pm 0.1$ & $2.7 \pm 0.1$ & $3.1 \pm 0.2$ & $2.4 \pm 0.2$ \\
     \hline
     \hline
     \multicolumn{5}{|c|}{Minimum $\langle H\rangle_{\Lambda_\text{S}}$ with hydrogen-like ansatz $R^{(0)}$}\\
     \hline
     \hline 
     $J^\pi$ & $\rho_1$ [GeV] & $\rho_2$ [GeV] & $\rho_3$ [GeV] & $E_0$ [GeV] \\
     \hline \hline
     $\frac{3}{2}^-$ & $0.4 \pm 0.1$ & $0.7 \pm 0.1$ & $1.2 \pm 0.4$ & $1.7 \pm 0.2$\\
     \hline
     $\frac{1}{2}^-$ & $0.4 \pm 0.1$ &  $0.5 \pm 0.1$ & $1.0 \pm 0.6$ & $1.8 \pm 0.2$\\
     \hline
     
\end{tabular}
\caption{ Optimal variational estimate for the candidate singlet $\Lambda_\text{S}$ mass in each $J^\pi$ channel, and the wavefunction variational parameters that deliver it. Our somewhat unexpected finding is that the $\frac{1}{2}^+$ candidate is above 2 GeV.}
\label{HydrogenLambdaS}
\end{table}


\section{Discussion}

It seems to us that the $qqq$ flavor singlets of lowest mass are well established to have quantum numbers $\frac{1}{2}^-$, $\frac{3}{2}^-$ and $\frac{1}{2}^+$, a result that we have rederived. 

Their masses are not too dissimilar in a harmonic-oscillator picture of baryons (the two negative parity states  would  basically be degenerate,  being in the first shell with $N=1$ excitation in a nonrelativistic setup and differing only in spin recoupling; the excitation energy of the positive parity state would be higher, jumping to the $N=2$ shell with positive parity). 

However, traditional flavor analysis~\cite{Guzey:2005rx} sometimes seem to ignore or do without the $\frac{1}{2}^+$ singlet, whose lowest mass candidate, $\Lambda(1810)$ as per the Graz effort~\cite{Melde:2008yr} has recently been put into question~\cite{Sarantsev:2019xxm} as unnecessary to explain scattering data. Experimentally reconfirming this state by different means then seems to be first-order business:
we have shown in an explicit calculation of the relativistic, chiral field-theory quark model extracted from Coulomb gauge QCD, and respecting its global symmetries, that this  $\frac{1}{2}^+$ singlet is heavier than the other two $J^\pi$ channels, and well above 2 GeV.

This means that even after accounting for mixing and for the effect of the nucleon-meson decay channels, it is unlikely to be in agreement with a 1.8 GeV mass.

The negative parity candidates, on the other hand, appear in the 1.7-1.8 GeV range (with a 0.1-0.2 GeV Monte Carlo error), consistently with expectations based on other quark approaches.

The two experimental $\Lambda$ candidates that could contain sizeable parts of this $qqq$ singlet wavefunction configuration appear in the 1.4-1.6 GeV range. This is 
expected from $\Delta M\sim \Gamma$ relating the real and imaginary shifts of the particle pole upon including open baryon-meson channels, and from mixing with color octet configurations.

That the radial-like excitation is heavier than the angular ones is not surprising upon reexamining the meson spectrum:
the $\rho(770)$ largely corresponds to the $q\overline{q}$ $(n_r=1)^3S_1$ state, the 
$(n_r=1)^3P_1$ corresponds to the $a_1(1260)$ and the $(n_r=2)^3S_1$ to the $\rho(1450)$. This entails the radial excitation to be 200 MeV above the orbital angular momentum one. A similar splitting separates the analogous $K_1(1270)$ and $K^*(1410)$.

In our baryon $\Lambda_\text{S}$ computation, and after allowing for the Monte Carlo uncertainties, it appears that the splitting is a factor of 2 larger. Whether this is \emph{(a)} an effect of the restriction to a flavor singlet, \emph{(b)} a variational effect (that we have not gotten a variational wave function close enough to the true one for the Hamiltonian in spite of the four families with two independent parameters tried), 
\emph{(c)} a model effect built into the Hamiltonian (in spite of its reasonable success in several other similar calculations), or \emph{(d)} a true feature of QCD (and possibly of nature) remains to be seen, but we have detected no obvious error that makes us suspect of the result in table~\ref{HydrogenLambdaS}.

If we were to dare a possible explanation, 
we would note the 150 MeV excess mass computed for the $\Delta$ baryon, that can decay strongly; a similar effect should be there for these hyperon resonances, and one would broadly expect it to grow with the particle mass as more decay channels (all ignored in a $qqq$ calculation) would be open.

Even after such effects are discounted, the fact that $qqq$ flavor-singlet baryon configurations have a mass so much larger than the ground state baryon octet is a consequence of the color degree of freedom, that forces them into an excited state.

Additionally to the mass, the issue of baryon-singlet identification can profit from studying decay-product distributions. We have examined them with reasonable EFT-based hadron models and have shown how the branching fractions depend on the hyperon mass. In those decays, $SU(3)$ symmetry is more easily extracted from data for decaying hyperons of higher mass: these see less pronounced effects of phase space and derivative couplings breaking $SU(3)$. 

Because those effects are rather violent for low-lying resonances, we hope that symmetry analysis of decays will be more useful to screen the $M>1.8$ GeV region for singlet candidates, where the experimental uncertainty can obfuscate the assignment much less.

Simultaneously, we hope to stimulate activity in lattice gauge theory towards untangling the octet-singlet $SU(3)$ flavor structure of the few low-lying resonances: it would be an interesting theoretical contribution to achieve such separation.

\acknowledgments
This project has received funding from the European Union's Horizon 2020 research and innovation programme under grant agreement No 824093 (STRONG-2020); and grants MINECO:FPA2016-75654-C2-1-P and MICINN:PID2019-108655GB-I00,  PID2019-106080GB-C21(Spain); Universidad Complutense de Madrid under research group 910309 and the IPARCOS institute.

\appendix

\section{Rarita-Schwinger spinors and $\Lambda_\text{S} (\frac{3}{2}^-)$ decay vertex} \label{ap:RS}

In this appendix we give some detail on the calculation of the two-body $\overline{|{\mathcal M}|^2}_{\Lambda_\text{S}\to B\Phi}$ for the spin-parity $\frac{3}{2}^-$ combination. 

Following~\cite{Rarita:1941mf}, we take a collection of four spin-$\frac{1}{2}$ Dirac spinors grouped as the components of a Minkowski-four vector, $\psi_{RS\ \mu}$ or simply $\psi_{\mu}$. Each of the spinors satisfies a free Dirac equation
\begin{equation}\label{DiracRS}
(i\not\!\partial - M_{\Lambda_\text{S}}) \psi_\mu(x)=0\ ,
\end{equation}
with $\partial_\mu \to -ip_\mu$ to convert to momentum eigenmodes.

Since $\psi_\nu$ is the tensor product of an object of spin $\frac{1}{2}$ (each Dirac spinor) and one that contains spins 0 and 1 (the four vector that collects them), this collection of spinors is not an irreducible representation of the rotation group (nor of the Lorentz one, of course) and contains two spin-$\frac{1}{2}$ representations in addition to the $3/2$ of interest to the decay at hand.

One of the unwanted representations is removed by imposing the condition (see for example~\cite{Hemmert:1997ye})
\begin{equation} \label{condRS1}
\gamma^\mu \psi_\mu = 0 \ ;
\end{equation}
as the four-vector index is contracted, this can be seen as a Dirac spinor condition.
A second such condition can be obtained~\cite{Milford:1955FJ} by multiplying
Eq.~(\ref{DiracRS}) by $\gamma^\mu$ and using Eq.~(\ref{condRS1}) to simplify,
\begin{equation}\label{condRS2}
\partial^\mu \psi_\mu = 0 \ .
\end{equation}
This removes the second unwanted spin $\frac{1}{2}$ representation, it being a condition in the $(0,1/2)$ representation of the Lorentz group cover.

To proceed quickly to $\Gamma_{\Lambda_\text{S}}$, we need the positive-spinor completeness relation equivalent to the Dirac spinor one
\begin{equation}\label{completeness}
\sum_\sigma u({\bf p},\sigma) \overline{u}({\bf p},\sigma) = \Lambda_+ = (\not\!p+M) \ .
\end{equation}
This will be a certain tensor
\begin{equation}
\sum_\sigma u^\mu({\bf p},\sigma) \overline{u}^\nu({\bf p},\sigma) = \Lambda^{\mu\nu}_+  = (\not\!p+M)\left( P^{3/2}\right)^{\mu\nu}
\end{equation}
with the spin-$\frac{1}{2}$ parts projected out.

To construct the tensor following~\cite{Siahaan}, let us first enforce Eq.~(\ref{condRS2}) by subtracting from the identity the projection over $p_\mu$,
\begin{equation}
\eta_\perp^{\mu\nu} := \eta^{\mu\nu}-\frac{p^\mu p^\nu}{p^2}\ .
\end{equation}

The resulting spinor $\eta_\perp^{\mu\nu}\psi_\nu$ obviously satisfies $p_\mu \eta_\perp^{\mu\nu}\psi_\nu=0$ and thus Eq.~(\ref{condRS2}), and falls in the reducible $(1,1/2)$ representation. It does not satisfy Eq.~(\ref{condRS1}) so we need to subtract another projection, 
forming
\begin{equation}\label{RSprojector}
\left(P^{3/2}\right)^{\mu\nu}= \eta_\perp^{\mu\nu} -\frac{1}{3p^2}
(p^\mu-\gamma^\mu \not\! p)
(p^\nu-\not\! p\gamma^\nu ) \ .
\end{equation}

It is easy to check several properties: first, 
\begin{equation}
\gamma_\mu\left(P^{3/2}\right)^{\mu\nu}= 0 = p_\mu \left(P^{3/2}\right)^{\mu\nu}
\end{equation}
and $\left(P^{3/2}\right)^{\mu\nu}\psi_\nu$ satisfies both Eq.~(\ref{condRS1}) and~(\ref{condRS2}).

Second, $P^{3/2}(p)$ commutes with $\not\! p$ so that it can be deployed to either side of 
$(\not\!p+M)$ in Eq.~(\ref{completeness}).
Third, it is indeed a projector,
\begin{equation}
\left(P^{3/2}\right)^\mu_\nu \left(P^{3/2}\right)^{\nu\rho}=\left(P^{3/2}\right)^{\mu\rho}
\end{equation}
so that only one copy appears in Eq.~(\ref{completeness}) that is constructed from two Rarita-Schwinger spinor collections.

Therefore, Eq.~(\ref{completeness}) and~(\ref{RSprojector}) suffice to reconstruct Eq.~(\ref{RStrace}) given the Lagrangian contribution yielding the decay. 

It remains to construct this decay potential, for which we once more take into account that chiral symmetry requires in leading order that the meson be derivatively coupled. We also need to analyze the parity. The positive component of the RS spinor collection, after Fourier transform, is a sum, with some Clebsch-Gordan coefficients, of a Dirac spinor $u$ multiplied by a spin-1 polarization vector $\epsilon_\nu$ and with a particle creation operator $b^\dagger$, namely
\begin{equation}
\psi^+_\nu\propto \int \sum ({\rm CG})\ \cdot \ u \ \cdot \ \epsilon_\nu\ \cdot\  b^\dagger\ .
\end{equation}

The creation operator carries the intrinsic parity of the $\Lambda_\text{S}$ particle $b^\dagger \arrowvert 0 \rangle$, which is $(-1)$ for the $\frac{3}{2}^-$ state; the spinor picks up a $\gamma_0$ in the Pauli-Dirac representation; and the vector $\epsilon_\nu$ changes sign under parity.
The $\gamma_0$ cancels out upon constructing a proper bilinear,
so we count the RS field as having parity opposite to that of the particle. 

Therefore, the parity-even effective vertex describing $\Lambda_\text{S}\to B\Phi$ is indeed that of Eq.~(\ref{decay2_3halves}).


\section{Integration limits in the Dalitz plane to integrate the three-body decays $1\to 2\ 3\ 4$}\label{sec:Dalitz}

In this paragraph we quickly sketch the three-body formalism needed to carry out three-body decay calculations such as those in subsec.~\ref{subsec:decay3}.
The independent invariant Dalitz variables chosen are
$m_{23}^2 \equiv p_{23}^2 \equiv (p_2 +p_3)^2$ and the analogous $m_{24}.$ 
Energy-momentum conservation $p_1=p_2+p_3+p_4$ entails that ${\bf p}_2$, ${\bf p}_3$ and ${\bf p}_4$ are coplanar in the cm system, where 
\begin{equation} \label{mDalitzofE}
m_{23}^2 = m_1^2 +m_4^2 -2m_1 E_4
\end{equation}
is rather simple, and once fixed, 
\begin{equation}
m_{24}^2 = m_2^2 +m_4^2 +2 (E_2E_4 - |\textbf{p}_2| |\textbf{p}_4|\cos \theta_{24})\ .
\end{equation}

The border of the physical region in the $(m_{23},m_{24})$ plane, the Dalitz plot, 
happens when the three-momenta are additionally collinear, $\cos \theta_{24}=1$.

First, let us give  the minimum values that the Dalitz variables can take; these are  $m_{ij}^{\text{min}} = m_i+m_j$, but they are not reached simultaneously. With a bit of work, inverting Eq.~(\ref{mDalitzofE}) for $E_4$ (and equivalently for $E_2$, $E_3$) to retrieve $E_i^{\text{min}}$, we obtain
\begin{equation}
m_{23}^2 (m_{24}^{\text{min}}) = m_1^2 +m_2^2 -\dfrac{m_2}{m_2+m_4} \bigl( m_1^2 -m_3^2 + (m_2 +m_4)^2\bigr)\ .
\end{equation}

Likewise, the maxima of each of the $ij$-Dalitz variables are reached when the remaining particle is left at rest, so that, for example,
\begin{equation}
m_{24}^{\text{max}} = m_1-m_3 , \qquad E_3 = m_3.
\end{equation}
Some algebra leads for example to 
\begin{equation}
E_4^{\text{max}} = \frac{(m_1-m_3)^2 +m_2^2 -m_4^2}{2(m_1-m_3)}
\end{equation}
and
\begin{equation}
m_{23}^2 (m_{42}^{\text{max}}) = m_1^2 +m_4^2 -2m_1 E_4 = m_4^2 +m_1m_3 -m_1 \dfrac{m_4^2 -m_2^2}{m_1-m_3}\ .
\end{equation}

The rest of the border can be obtained from the collinearity condition, the on-shell and momentum conservation conditions, yielding for example
\begin{eqnarray} 
E_4^\pm = \frac{1}{2m_{24}^2}\left[ \frac{(m_{24}^2 +m_4^2-m_2^2) (m_1^2 +m_{24}^2 -m_3^2)}{2m_1} \right. \nonumber \\
\pm \sqrt{\left(\frac{m_1^2 +m_{24}^2 -m_3^2}{2m_1}\right)^2-m_{24}^2} \\ \nonumber
\times \left.  \sqrt{ (m_{24}^2 +m_4^2-m_2^2)^2 -4m_{24}^2 m_4^2 }\right] 
\end{eqnarray}
that can be substituted into
\be
m_{23}^2 (m_{24})_\pm = m_1^2 +m_4^2 -2 m_1 E_4^\pm (m_{24}^2) 
\ee
to complete the figure in the Dalitz plane. With the computed borders, the three-body widths are then straight-forward to extract,
\ba
\left.\Gamma (m_1) \right|_{m_2,m_3,m_4}= \ \frac{1}{32 (2\pi)^3 m_1^3}\times  \\ \nonumber 
 \int_{(m_{24}^{\rm min})^2}^{(m_{24}^{\rm max})^2} \text{d}m_{24}^2 \int_{m_{23}^2 (m_{24})_+}^{m_{23}^2 (m_{24})_-} \text{d}m_{23}^2 |{\mathcal M}(m_{23}^2, m_{24}^2)|^2 \ .
\ea



\begin{thebibliography}{100}
 
\bibitem{Perl:2009zz}
M.~L.~Perl, E.~R.~Lee and D.~Loomba,
Ann. Rev. Nucl. Part. Sci. \textbf{59}, 47-65 (2009)
doi:10.1146/annurev-nucl-121908-122035

\bibitem{Bergsma:1984yn}
F.~Bergsma \textit{et al.} [CHARM],
Z. Phys. C \textbf{24}, 217 (1984)
doi:10.1007/BF01410361


\bibitem{HidalgoDuque:2011je}
R.~L.~Delgado, C.~Hidalgo-Duque and F.~J.~Llanes-Estrada,
Few Body Syst. \textbf{54}, 1705-1717 (2013)
doi:10.1007/s00601-012-0500-5. 


\bibitem{Close:1979bt}
F.~E.~Close,
``An Introduction to Quarks and Partons,''
Academic Press, London (1980)
ISBN 012175152X

\bibitem{Halzen:1984mc}
F.~Halzen and A.~D.~Martin,
``QUARKS AND LEPTONS: AN INTRODUCTORY COURSE IN MODERN PARTICLE PHYSICS,''
John Wiley \& sons, Hoboken NJ; 1st edition (1984)
ISBN: 0471887412.


\bibitem{Jido:2003cb}
D.~Jido, J.~A.~Oller, E.~Oset, A.~Ramos and U.~G.~Meissner,
Nucl. Phys. A \textbf{725} (2003), 181-200
doi:10.1016/S0375-9474(03)01598-7.




\bibitem{Meissner:2020khl}
U.~G.~Meißner,
Symmetry \textbf{12} (2020), 981
doi:10.3390/sym12060981.



\bibitem{Engel:2013lea}
G.~P.~Engel {\it et al.}
PoS \textbf{Hadron2013}, 118 (2013)
doi:10.22323/1.205.0118
[arXiv:1311.6579 [hep-ph]].

\bibitem{Hall:2014uca}
J.~M.~M.~Hall {\it et al.}
Phys. Rev. Lett. \textbf{114} (2015), 132002
doi:10.1103/PhysRevLett.114.132002.


\bibitem{Melde:2008yr} 
  T.~Melde, W.~Plessas and B.~Sengl,
  Phys.\ Rev.\ D {\bf 77}, 114002 (2008)
  doi:10.1103/PhysRevD.77.114002.
  
  

\bibitem{Pauli:2019ydi}
P.~Pauli [GlueX],  procs. Int. Nuclear Physics Conference 2019, 
[arXiv:1909.10877 [nucl-ex]].


  
\bibitem{Lin:2011ti}
H.~W.~Lin,
Chin. J. Phys. \textbf{49}, 827 (2011).

\bibitem{Lin:2008rb}
H.~W.~Lin,
Nucl. Phys. B Proc. Suppl. \textbf{187}, 200-207 (2009)
doi:10.1016/j.nuclphysbps.2009.01.029.


\bibitem{Guzey:2005rx}
V.~Guzey and M.~V.~Polyakov,
Annalen Phys. \textbf{13} (2004), 673-681
doi:10.1002/andp.200410109;
{\it ibid.} ``SU(3) systematization of baryons,'' preprint
[arXiv:hep-ph/0512355 [hep-ph]].


\bibitem{Sarantsev:2019xxm}
A.~V.~Sarantsev {\it et al.}
Eur. Phys. J. A \textbf{55}, 180 (2019)
doi:10.1140/epja/i2019-12880-5




\bibitem{Qin:2019hgk}
S.~x.~Qin, C.~D.~Roberts and S.~M.~Schmidt,
Few Body Syst. \textbf{60}, 26 (2019)
doi:10.1007/s00601-019-1488-x



\bibitem{Ronniger:2011td}
M.~Ronniger and B.~C.~Metsch,
Eur. Phys. J. A \textbf{47}, 162 (2011)
doi:10.1140/epja/i2011-11162-8




\bibitem{Loring:2001kx}
U.~Loring, B.~C.~Metsch and H.~R.~Petry,
Eur. Phys. J. A \textbf{10}, 395-446 (2001)
doi:10.1007/s100500170105



\bibitem{Nakajima:2001js}
N.~Nakajima, H.~Matsufuru, Y.~Nemoto and H.~Suganuma,
AIP Conf. Proc. \textbf{594} (2001), 349
doi:10.1063/1.1425521. 






\bibitem{Azizi:2020ljx}
K.~Azizi, Y.~Sarac and H.~Sundu,
Phys. Rev. D \textbf{102} (2020), 034007
doi:10.1103/PhysRevD.102.034007.

\bibitem{Ebert:2011kk}
D.~Ebert, R.~N.~Faustov and V.~O.~Galkin,
Phys. Rev. D \textbf{84} (2011), 014025
doi:10.1103/PhysRevD.84.014025




\bibitem{Rosner:2006jz}
J.~L.~Rosner,
J. Phys. G \textbf{34}, S127-S148 (2007)
doi:10.1088/0954-3899/34/7/S07



\bibitem{Ruijl:2017dtg}
B.~Ruijl, T.~Ueda and J.~Vermaseren,
[arXiv:1707.06453 [hep-ph]].



\bibitem{Thomas:2001}
S.K. Lin and W. Weise, Molecules  {\bf 6}(12):1041–1043 (2001)  doi:10.3390/61201041;
Anthony W. Thomas, Wolfram Weise,
``The Structure of the Nucleon'' 
Wiley-VCH, Berlin, 2001.  ISBN: 3-527-40297-7. 


\bibitem{Review:2016}
Review of Particle Physics, Chinese Physics C, vol. 40, No. 10 (2016), p.1578;
P.~A.~Zyla \textit{et al.} [Particle Data Group],
PTEP \textbf{2020}, 083C01 (2020)
doi:10.1093/ptep/ptaa104


  


\bibitem{Robertson:1998va}
D.~G.~Robertson {\it et al.}
Phys. Rev. D \textbf{59}, 074019 (1999)
doi:10.1103/PhysRevD.59.074019 .

\bibitem{LlanesEstrada:1999uh}
F.~J.~Llanes-Estrada and S.~R.~Cotanch,
Phys. Rev. Lett. \textbf{84}, 1102-1105 (2000)
doi:10.1103/PhysRevLett.84.1102

\bibitem{Szczepaniak:2001rg}
A.~P.~Szczepaniak and E.~S.~Swanson,
Phys.\ Rev.\ D \textbf{65}, 025012 (2002)
doi:10.1103/PhysRevD.65.025012 .


\bibitem{Eichten:1979ms}
E.~Eichten, K.~Gottfried, T.~Kinoshita, K.~D.~Lane and T.~M.~Yan,
Phys. Rev. D \textbf{21}, 203 (1980)
doi:10.1103/PhysRevD.21.203



\bibitem{LlanesEstrada:2000jw}
F.~J.~Llanes-Estrada {\it et al.}
Nucl. Phys. A \textbf{710}, 45-54 (2002)
doi:10.1016/S0375-9474(02)01090-4

\bibitem{LlanesEstrada:2005jf}
F.~J.~Llanes-Estrada, P.~Bicudo and S.~R.~Cotanch,
Phys. Rev. Lett. \textbf{96}, 081601 (2006)
doi:10.1103/PhysRevLett.96.081601 .

\bibitem{LlanesEstrada:2000hj}
F.~J.~Llanes-Estrada and S.~R.~Cotanch,
Phys. Lett. B \textbf{504}, 15-20 (2001)
doi:10.1016/S0370-2693(01)00290-8

\bibitem{General:2006ed}
I.~J.~General, S.~R.~Cotanch and F.~J.~Llanes-Estrada,
Eur. Phys. J. C \textbf{51}, 347-358 (2007)
doi:10.1140/epjc/s10052-007-0298-3


\bibitem{TorresRincon:2010fu}
J.~M.~Torres-Rincon and F.~J.~Llanes-Estrada,
Phys. Rev. Lett. \textbf{105}, 022003 (2010)
doi:10.1103/PhysRevLett.105.022003


\bibitem{Wang:2008mw}
P.~Wang, S.~R.~Cotanch and I.~J.~General,
Eur. Phys. J. C \textbf{55}, 409-415 (2008)
doi:10.1140/epjc/s10052-008-0605-7



\bibitem{General:2007bk}
I.~J.~General {\it et al.}
Phys. Lett. B \textbf{653}, 216-223 (2007)
doi:10.1016/j.physletb.2007.08.015


\bibitem{Abreu:2019adi}
L.~M.~Abreu {\it et al.}
Phys. Rev. D \textbf{100}, no.11, 116012 (2019)
doi:10.1103/PhysRevD.100.116012
[arXiv:1908.11154 [hep-ph]].

\bibitem{Abreu:2020ttf}
L.~M.~Abreu, F.~M.~d.~Júnior and A.~G.~Favero,
[arXiv:2007.07849 [hep-ph]].

\bibitem{Abreu:2020wio}
L.~M.~Abreu, F.~M.~da Costa Júnior and A.~G.~Favero,
Phys. Rev. D \textbf{101}, no.11, 116016 (2020)
doi:10.1103/PhysRevD.101.116016


\bibitem{Alkofer:2018yjm}
See for recent work {\it e.g.}
R.~Alkofer, C.~S.~Fischer and H.~Sanchis-Alepuz,
EPJ Web Conf. \textbf{181}, 01013 (2018)
doi:10.1051/epjconf/201818101013
[arXiv:1802.09775 [hep-ph]];
Z.~F.~Cui {\it et al.}
[arXiv:2003.11655 [hep-ph]].


\bibitem{Choi:2017uos}
H.~M.~Choi and C.~R.~Ji,
Phys. Rev. D \textbf{95}, 056002 (2017)
doi:10.1103/PhysRevD.95.056002
 
\bibitem{Gomez-Rocha:2013zma}
M.~Gomez-Rocha, W.~Schweiger and O.~Senekowitsch,
Few Body Syst. \textbf{55}, 697-700 (2014)
doi:10.1007/s00601-013-0779-x



\bibitem{LlanesEstrada:2004wr}
  F.~J.~Llanes-Estrada {\it et al.}
  Phys.\ Rev.\ C \textbf{70}, 035202 (2004)
  doi:10.1103/PhysRevC.70.035202.




\bibitem{Hahn:2004fe.Cuba}
T.~Hahn,
Comput. Phys. Commun. \textbf{168}, 78-95 (2005)
doi:10.1016/j.cpc.2005.01.010 .




\bibitem{Lepage:1980dq}
G.~P.~Lepage,
``VEGAS: AN ADAPTIVE MULTIDIMENSIONAL INTEGRATION PROGRAM,''
preprint CLNS-80/447.



\bibitem{Bicudo:1989si}
P.~J.~d.~A.~Bicudo and J.~E.~F.~T.~Ribeiro,
Phys. Rev. D \textbf{42} (1990), 1625-1634
doi:10.1103/PhysRevD.42.1625

\bibitem{LeYaouanc:1984ntu}
A.~Le Yaouanc, L.~Oliver, S.~Ono, O.~Pene and J.~C.~Raynal,
Phys. Rev. D \textbf{31} (1985), 137-159
doi:10.1103/PhysRevD.31.137



\bibitem{Bicudo:2009cr}
P.~Bicudo {\it et al.}
Phys. Rev. Lett. \textbf{103}, 092003 (2009)
doi:10.1103/PhysRevLett.103.092003.



\bibitem{Bicudo:2016eeu}
P.~Bicudo {\it et al.}
Phys. Rev. D \textbf{94}, 054006 (2016)
doi:10.1103/PhysRevD.94.054006 .

\bibitem{LlanesEstrada:2011jd}
F.~J.~Llanes-Estrada and G.~M.~Navarro,
Mod. Phys. Lett. A \textbf{27}, 1250033 (2012)
doi:10.1142/S0217732312500332 .



\bibitem{Rarita:1941mf}
W.~Rarita and J.~Schwinger,
Phys. Rev. \textbf{60}, 61 (1941)
doi:10.1103/PhysRev.60.61



\bibitem{Hemmert:1997ye}
T.~R.~Hemmert, B.~R.~Holstein and J.~Kambor,
J. Phys. G \textbf{24}, 1831-1859 (1998)
doi:10.1088/0954-3899/24/10/003 .

\bibitem{Milford:1955FJ}
F.~J.~Milford, Phys. Rev. {\bf 98}, 1488 (1955).

\bibitem{Siahaan}
H.~M.~Siahaan, thesis presented to the Technological Institute of Bandung, available in {\tt http://www.fisikanet.lipi.go.id/data\\ /1014224401/data/1202810210.pdf}.


\end{thebibliography}
\end{document}